\documentclass[twoside,twocolumn,9pt]{article}
\usepackage{extsizes}
\usepackage[super,sort&compress,comma]{natbib} 
\usepackage[version=3]{mhchem}
\usepackage[left=1.5cm, right=1.5cm, top=1.785cm, bottom=2.0cm]{geometry}
\usepackage{balance}
\usepackage{mathptmx}
\usepackage{sectsty}
\usepackage{graphicx} 
\usepackage{lastpage}
\usepackage[format=plain,justification=justified,singlelinecheck=false,font={stretch=1.125,small,sf},labelfont=bf,labelsep=space]{caption}
\usepackage{float}
\usepackage{fancyhdr}
\usepackage{fnpos}
\usepackage[english]{babel}
\addto{\captionsenglish}{%
  
}
\usepackage{array}
\usepackage{droidsans}
\usepackage{charter}
\usepackage[T1]{fontenc}
\usepackage[usenames,dvipsnames]{xcolor}
\usepackage{setspace}
\usepackage[compact]{titlesec}
\usepackage[colorlinks=true]{hyperref}
\usepackage{multirow}
\usepackage{authblk}
\usepackage{afterpage}
\usepackage{epstopdf}

\usepackage{epstopdf}

\definecolor{cream}{RGB}{222,217,201}

\begin{document}

\pagestyle{fancy}
\thispagestyle{plain}
\fancypagestyle{plain}{
\renewcommand{\headrulewidth}{0pt}
}

\makeFNbottom
\makeatletter
\renewcommand\LARGE{\@setfontsize\LARGE{15pt}{17}}
\renewcommand\Large{\@setfontsize\Large{12pt}{14}}
\renewcommand\large{\@setfontsize\large{10pt}{12}}
\renewcommand\footnotesize{\@setfontsize\footnotesize{7pt}{10}}
\makeatother

\renewcommand{\thefootnote}{\fnsymbol{footnote}}
\renewcommand\footnoterule{\vspace*{1pt}%
\color{cream}\hrule width 3.5in height 0.4pt \color{black}\vspace*{5pt}} 
\setcounter{secnumdepth}{5}

\makeatletter 
\renewcommand\@biblabel[1]{#1}            
\renewcommand\@makefntext[1]%
{\noindent\makebox[0pt][r]{\@thefnmark\,}#1}
\makeatother 
\renewcommand{\figurename}{\small{Fig.}~}
\sectionfont{\sffamily\Large}
\subsectionfont{\normalsize}
\subsubsectionfont{\bf}
\setstretch{1.125} 
\setlength{\skip\footins}{0.8cm}
\setlength{\footnotesep}{0.25cm}
\setlength{\jot}{10pt}
\titlespacing*{\section}{0pt}{4pt}{4pt}
\titlespacing*{\subsection}{0pt}{15pt}{1pt}

\fancyfoot{}
\fancyfoot[LO,RE]{\vspace{-7.1pt}\includegraphics[height=9pt]{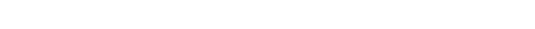}}
\fancyfoot[CO]{\vspace{-7.1pt}\hspace{13.2cm}\includegraphics{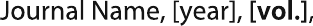}}
\fancyfoot[CE]{\vspace{-7.2pt}\hspace{-14.2cm}\includegraphics{head_foot/RF}}
\fancyfoot[RO]{\footnotesize{\sffamily{1--\pageref{LastPage} ~\textbar  \hspace{2pt}\thepage}}}
\fancyfoot[LE]{\footnotesize{\sffamily{\thepage~\textbar\hspace{3.45cm} 1--\pageref{LastPage}}}}
\fancyhead{}
\renewcommand{\headrulewidth}{0pt} 
\renewcommand{\footrulewidth}{0pt}
\setlength{\arrayrulewidth}{1pt}
\setlength{\columnsep}{6.5mm}
\setlength\bibsep{1pt}

\makeatletter 
\newlength{\figrulesep} 
\setlength{\figrulesep}{0.5\textfloatsep} 

\newcommand{\topfigrule}{\vspace*{-1pt}%
\noindent{\color{cream}\rule[-\figrulesep]{\columnwidth}{1.5pt}} }

\newcommand{\botfigrule}{\vspace*{-2pt}%
\noindent{\color{cream}\rule[\figrulesep]{\columnwidth}{1.5pt}} }

\newcommand{\dblfigrule}{\vspace*{-1pt}%
\noindent{\color{cream}\rule[-\figrulesep]{\textwidth}{1.5pt}} }

\makeatother

\twocolumn[
  \begin{@twocolumnfalse}
{\includegraphics[height=30pt]{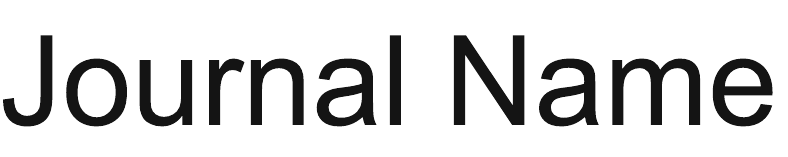}\hfill\raisebox{0pt}[0pt][0pt]{\includegraphics[height=55pt]{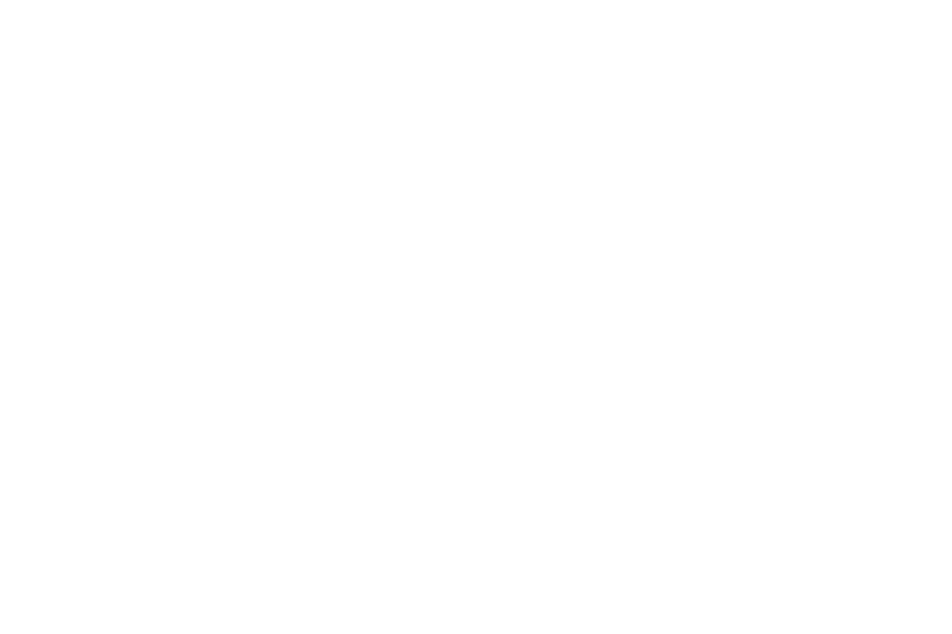}}\\[1ex]
\includegraphics[width=18.5cm]{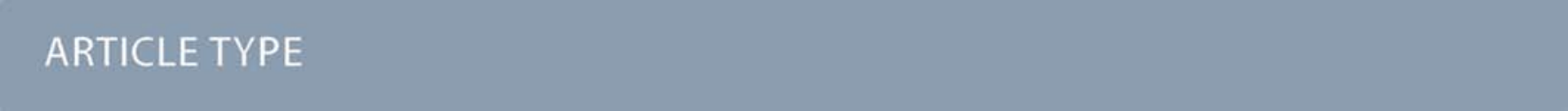}}\par
\vspace{1em}
\sffamily
\begin{tabular}{m{4.5cm} p{13.5cm} }

\includegraphics{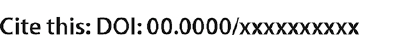} & \noindent\LARGE{\textbf{The three-center two-positron bond}} \\
\vspace{0.3cm} & \vspace{0.3cm} \\

 & \noindent\large{Jorge Charry\textit{$^{a}$}, F\'elix Moncada\textit{$^{b,e}$}, Matteo Barborini\textit{$^{a}$}, Laura Pedraza-Gonz{\'a}lez\textit{$^{c}$}, M\'arcio T. do N. Varella\textit{$^{d}$}, 
 Alexandre Tkatchenko\textit{$^{a}$}, 
 Andres Reyes\textit{$^{e}$} }
 \\

\includegraphics{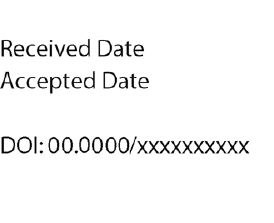} & \noindent\normalsize{Computational studies have shown that one or more positrons can stabilize two repelling atomic anions through the formation of two-center positronic bonds. In the present work, we study the energetic stability of a system containing two positrons and three hydride anions, namely 2\ce{e^+[H^{3-}_3]}. 
To this aim, we performed a preliminary scan of the potential energy surface of the system with both electrons and positron in a spin singlet state, with a multi-component MP2 method, that was 
further refined with variational and diffusion Monte Carlo calculations, and confirmed an equilibrium geometry with \ce{D_${3h}$} symmetry. The local stability of 2\ce{e^+[H^{3-}_3]} is demonstrated by analyzing the vertical detachment and adiabatic energy dissociation channels. Bonding properties of the positronic compound, such as the equilibrium interatomic distances, force constants, dissociation energies, and bonding densities are compared with those of the purely electronic \ce{H^+_3} and \ce{Li^+_3} systems. 
Through this analysis, we find compelling similarities between the 2\ce{e^+[H^{3-}_3]} compound and the trilithium cation.
Our results strongly point out the formation of a non-electronic three-center two-positron bond, analogous to the well known three-center two-electron counterparts, which is fundamentally distinct from the two-center two-positron bond [D. Bressanini, \textit{J. Chem. Phys.} \textbf{155}, 054306 (2021)], thus extending the concept of positron bonded molecules. 
} \\

\end{tabular}

 \end{@twocolumnfalse} \vspace{0.6cm}

  ]

\renewcommand*\rmdefault{bch}\normalfont\upshape
\rmfamily
\section*{}
\vspace{-1cm}

\footnotetext{\textit{$^{a}$~Department of Physics and Materials Science, University of Luxembourg, L-1511 Luxembourg City, Luxembourg.}}
\footnotetext{\textit{$^{b}$~Department of Physics, AlbaNova University Center, Stockholm University, S-106 91 Stockholm, Sweden.}}
\footnotetext{\textit{$^{c}$~Department of Chemistry and Industrial Chemistry, University of Pisa, Via Moruzzi 13, 56124, Pisa, Italy. }}
\footnotetext{\textit{$^{d}$~Instituto de Física, Universidade de São Paulo, Rua do Matão 1731, São Paulo, São Paulo, 05508-090, Brazil. }}
\footnotetext{\textit{$^{e}$~Department of Chemistry, Universidad Nacional de Colombia, Av. Cra 30\#45–03, Bogot\'a, Colombia; E-mail: areyesv@unal.edu.co}}

\footnotetext{\dag~Electronic Supplementary Information (ESI) available: 
Total energies of the one, two and three atomic species employed in the dissociation energies calculations.
Quantum Monte Carlo energies computed at each point of the potential energy curves.}



\section{Introduction}

The techniques to accumulate and manipulate positrons \cite{fajans2020plasma} and positronium (Ps) atoms \cite{cassidy2018experimental} at very low energies have opened up new research areas in molecular sciences. 
For instance, a great variety of positronic molecules have been produced and detected \cite{danielson2015plasma}. In these experiments, positron attachment to vibrationally excited molecules gives rise to positronic compounds. 
The existence of a different kind of positronic molecules was predicted by Charry et al. \cite{charry2018binding} based on computational simulations. Two otherwise repelling hydride anions can be bound by a positron, and the analysis of the positronic orbital suggested the formation of a positron covalent bond. 
Subsequent computational studies conducted by Moncada et al. \cite{moncada2020covalent} pointed out that similar bonds should also exist in positronic dihalides.

The positronic dihydride system, denoted as \ce{e^+[H^{2-}_2]}, has attracted considerable attention. Goli and Shahbazian \cite{Goli2019} addressed the nature of the interaction employing Multi-Component Quantum Theory of Atoms in Molecules (MC-QTAIM). The authors concluded that the accumulation of positron density between the two nuclei would underlie the bonding process, acting as ''a positronic glue'' that converts the repulsive \ce{H^{2-}_2} system into a stable one.
The energy stability of positron-bonded dihydrides was further confirmed by quantum Monte Carlo (QMC) calculations performed by Ito et al. \cite{Ito2020}, Bressanini \cite{Bressanini2021} and Charry et al.\cite{cha+22jctc}. 
Bressanini also concluded that the bonding is not limited to one positron and in later works\cite{Bressanini2021-2} he showed that two singlet-coupled positrons can form a bond between two hydride anions, thus giving rise to the 2\ce{e^{+}[H^{2-}_2]} molecule, or simply \ce{(PsH)_2}.
The equilibrium internuclear distance was reduced from 6.4 bohr to 6.0 bohr as the number of positrons was increased from 1 to 2. Unexpectedly, the bond length shortening was accompanied by a reduction in the bond energy (BE), from 23.5 m$E_\text{h}$ to 10.4 m$E_\text{h}$.

These fascinating theoretical results raise questions on how complex positron-bonded compounds can become, or on how far the analogy between electronic and positronic bonds can be stretched.
Along these lines, Bressanini \cite{3e+} recently reported the local energy stability of 3\ce{e^{+}[H^{2-}_2]}, or more simply \ce{e^{+}[PsH]_2}, a stable molecular system containing three positrons, which can be understood as resulting from the reaction \ce{[PsH]_2 + e+}. The present study further explores positronic bonding mechanisms beyond the previously studied two-center to three-center systems. 
Based on high-level QMC simulations, we demonstrate that the system comprising three hydride anions and two positrons, denoted as 2\ce{e^+[H^{3-}_3]}, is energetically stable. 
Our results, obtained for the singlet state in the \ce{D$_{3h}$} symmetry, point out the existence of a three-center two-positron (3c2p) bond.
To gain further insight into the 3c2p bond, we compare it against the analogous three-center two-electron bond 3c2e, which is a well established bonding mechanism found in purely electronic molecules\cite{lobayan2011electronic, green2012occurrence, smirnova2013hexanuclear, duchimaza2018quasi}. 

The \ce{H^+_3} molecule is probably the best known example of 3c2e bonds \cite{chuluunbaatar2020D3h}, in view of its astrochemical relevance \cite{herbst2000astrochemistry,miller2000role,oka2006interstellar}.
While the trihydrogen cation could be rated the closest analogue of 2\ce{e^+[H^{3-}_3]} at first glance, previous studies have consistently shown that the properties of electronic and positronic bonds are closest when the systems have isoelectronic ion cores \cite{charry2018binding,moncada2020covalent}. 
We therefore consider \ce{Li_3^+} as an alternative purely electronic analogue of the di-positronic tri-hydride compound for comparison of the bonding properties. 
Our results confirm that 2\ce{e^+[H^{3-}_3]} is stable against dissociation into chemically meaningful products.
The presently reported results, which indicate the existence of a new kind of di-positronic stable system, may hopefully expand the landscape of the field.

This paper is organized as follows: In Sec. \ref{methods} the methods and numerical procedures are summarized; the properties of the di-positronic system and the purely electronic analogues are presented and discussed in Sec.  \ref{results}; concluding remarks and perspectives for future work are outlined in Sec. \ref{conclusions}.

\section{Methods}\label{methods}
This study considers electrons and positrons as quantum particles and atomic nuclei as point charges within the Born-Oppenheimer approximation. Exploratory studies of the potential energy surface (PES) of the 2\ce{e^+[H^{3-}_3]} system were performed at the MP2 level of theory.
The MP2 expressions for systems composed of electrons and positrons can be found in Refs. \citenum{Charry2014,reyes2018}.   
Equilibrium energies and geometries were further refined with the variational Monte Carlo (VMC) and diffusion Monte Carlo (DMC) methods.

\subsection{Quantum Monte Carlo methods}

QMC methods are stochastic techniques employed to integrate the time independent Schr\"odinger equation for a given trial wave function, the most common of which are VMC and DMC).\cite{fou+01rmp,kal+08ch8,bec+17}
The stochastic integration in VMC is carried out with the Metropolis-Hastings algorithm\cite{met+53jcp,has+70b}. The energy functional is evaluated
over a trial wave function which is written as the product of a fermionic part, such as a Slater determinant, and a bosonic component, the Jastrow factor, used to describe explicit correlation between particles.
The trial wave function used in this study of electron-positron systems is written as the product 
\begin{equation}
\Psi = \det \left [ \textbf{S}_{e\uparrow} \right ] \det \left [ \textbf{S}_{e\downarrow} \right ] \det \left [ \textbf{S}_{p\uparrow} \right ] \det \left [ \textbf{S}_{p\downarrow} \right ] e^{\mathcal{J}(\bar{\textbf{r}}^e,\textbf{r}^p; \bar{\textbf{R}})}, 
\label{VMCfunction}
\end{equation}
where $\textbf{S}_{e\uparrow}$ and $\textbf{S}_{e\downarrow}$ are the electronic Slater matrices associated to the two spin populations, $\textbf{S}_{p\uparrow}$ and $\textbf{S}_{p\downarrow}$ are the corresponding matrices for the positrons, and $e^{\mathcal{J}}$ is the Jastrow factor. 
Several QMC studies have employed similar trial wave functions \cite{Yoshida1996,Kita2009,Yamada2014,Yamada2014a,Ito2020,cha+22jctc} based on products of determinants constructed via atomic basis sets. This particular Jastrow factor was introduced in Ref. \citenum{cha+22jctc} and it is now generalized for two-positron systems in the present work.
The Jastrow factor is built as the sum 
\begin{multline}
\mathcal{J}(\bar{\textbf{r}}^e,\bar{\textbf{r}}^p; \bar{\textbf{R}}) = \mathcal{J}_c^{en}(\bar{\textbf{r}}^e,\bar{\textbf{R}})
+\mathcal{J}_c^{pn}(\bar{\textbf{r}}^p,\bar{\textbf{R}}) 
+\mathcal{J}_c^{ee}(\bar{\textbf{r}}^e) + \\
+\mathcal{J}_c^{pp}(\bar{\textbf{r}}^p)
+\mathcal{J}_c^{ep}(\bar{\textbf{r}}^e,\bar{\textbf{r}}^p)
+\mathcal{J}_{3/4}(\bar{\textbf{r}}^e,\bar{\textbf{r}}^p; \bar{\textbf{R}})
\end{multline}
of five functions describing respectively the electron-nucleus ($\mathcal{J}_c^{en}(\bar{\textbf{r}}^e,\bar{\textbf{R}})$), positron-nucleus ($\mathcal{J}_c^{pn}(\bar{\textbf{r}}^p,\bar{\textbf{R}})$), electron-electron ($\mathcal{J}_c^{ee}(\bar{\textbf{r}}^e)$), positron-positron ($\mathcal{J}_c^{pp}(\bar{\textbf{r}}^p)$) and electron-positron ($\mathcal{J}_c^{ep}(\bar{\textbf{r}}^e,\bar{\textbf{r}}^p)$) cusps, and a term that describes the dynamical correlation between the fermionic particles in the field of the nuclei ($\mathcal{J}_{3/4}(\bar{\textbf{r}}^e,\textbf{r}^p; \bar{\textbf{R}})$), which is an extension of the one defined in ref. \citenum{Casula2003}. 
The functions used to describe the cusp conditions are of two types. For particle pairs with the same charge we employ slowly decaying functions\cite{Fahy1990} together with a linear combination of Gaussian-type functions (GTFs),
\begin{multline}
\mathcal{J}_c^{ee}(\bar{\textbf{r}}^e) = \sum_{i>j=1}^{N_e} \left( -\frac{\Gamma}{\beta^{ee}_{0}(1+\beta^{ee}_{0}|\mathbf{r}^e_{i}-\mathbf{r}^e_{j}|)}
    \right . +\\ + \left . \sum_{g=1}^{G} \beta^{ee}_{g}  e^{ -\zeta^{ee}_{g} |\mathbf{r}^e_{i}-\mathbf{r}^e_{j}|^2 }
    \right)
\end{multline}
\begin{multline}
  \mathcal{J}_c^{pp}(\bar{\textbf{r}}^p) = \sum_{i>j=1}^{N_p} \left( -\frac{\Gamma}{\beta^{pp}_{0}(1+\beta^{pp}_{0}|\mathbf{r}^p_{i}-\mathbf{r}^p_{j}|)}
    \right . + \\ + \left . \sum_{g=1}^{G}\beta^{pp}_{g}  e^{  -\zeta^{pp}_{g} |\mathbf{r}^p_{i}-\mathbf{r}^p_{j}|^2 }
    \right)
\end{multline}
\begin{multline}
\mathcal{J}_c^{pn}(\bar{\textbf{r}}^p,\bar{\textbf{R}}) = \sum_{i=1}^{N_p} \sum_{I=1}^{N_n} \left( -\frac{Z_I}{\beta^{pI}_{0}(1+\beta^{pI}_{0}|\mathbf{r}^{p}_i-\mathbf{R}_{I}|)}
    \right . + \\ + \left . \sum_{g=1}^{G} \beta^{pI}_{g}  e^{  -\zeta^{pI}_{g} |\mathbf{r}^{p}_i-\mathbf{R}_{I}|^2 }
    \right)
\end{multline}
while for particles with opposite charges we employ faster decaying cusp functions\cite{Becker1968}, along with the linear combination of GTFs,
\begin{equation}
\mathcal{J}_c^{en}(\bar{\textbf{r}}^e,\bar{\textbf{R}}) = \sum_{i=1}^{N_e}  \sum_{I=1}^{N_n} \left( \frac{Z_I}{\beta^{eI}_{0}}e^{-\beta^{eI}_{0}|\mathbf{r}^e_{i}-\mathbf{R}_{I}|}
    + \sum_{g=1}^{G} \beta^{eI}_{g} e^{  -\zeta^{eI}_{g} |\mathbf{r}^e_{i}-\mathbf{R}_{I}|^2}
    \right)
\end{equation}
\begin{equation}
  \mathcal{J}_c^{ep}(\bar{\textbf{r}}^e,\bar{\textbf{r}}^p) = \sum_{j=1}^{N_p} \sum_{i=1}^{N_e} \left( \frac{\Gamma}{\beta^{ep}_{0}}e^{-\beta^{ep}_{0}|\mathbf{r}^e_{i}-\mathbf{r}^p_{j}|}
    + \sum_{g=1}^{G} \beta^{ep}_{g} e^{ -\zeta^{ep}_{g} |\mathbf{r}^e_{i}-\mathbf{r}^p_{j}|^2}
    \right)
\end{equation}
Here $\beta$ and $\zeta$ are sets of variational parameters, $Z_I$ are the nuclear charges and $\Gamma$ is a constant that is equal to $1/2$ and $1/4$ respectively for distinguishable and indistinguishable particles to correctly describe the cusp conditions. In this investigation, the number of GTFs used in the cusps expansions is $G=5$. To reduce the variance in the wave function, the variational parameters for distinguishable or indistinguishable electronic and positronic pairs are optimized separately, also reducing the eventual spin contamination.\cite{hua+98jcp}

Finally, the dynamical Jastrow factor\cite{Casula2003,cha+22jctc} is written as a linear combination of products of non-normalized atomic orbitals $\chi_{q}$,
\begin{multline}
\mathcal{J}_{3/4}(\bar{\textbf{r}}^e,\bar{\textbf{r}}^p; \bar{\textbf{R}}) =
\sum_{j>i=1}^{N_e} \sum_{q,p=1}^{Q} \gamma_{qp} \chi_q(\textbf{r}^e_i) \chi_p(\textbf{r}^e_j) + \\
+\sum_{j>i=1}^{N_p} \sum_{q,p=1}^{Q} \mu_{qp} \chi_q(\textbf{r}^p_i) \chi_p(\textbf{r}^p_j) 
+ \sum_{i=1}^{N_e}\sum_{j=1}^{N_p}  \sum_{q,p=1}^{Q} \nu_{qp} \chi_q(\textbf{r}^e_i) \chi_p(\textbf{r}^p_j),
\end{multline}
where $Q$ is the total length of the basis set, while $\gamma_{qp}$,  $\mu_{qp}$ and $\nu_{qp}$ are coupling parameters that are fully optimized.
To further avoid the spin contamination introduced by the Jastrow factor, the $\gamma_{qp}$ and $\mu_{qp}$ parameters, which respectively describe correlation between electron pairs and positron pairs, are symmetric ($\gamma_{qp}=\gamma_{pq}$ and $\mu_{qp}=\mu_{pq}$). In this investigation, (3s2p1d) uncontracted GTFs were employed as the $\chi_{q}$ orbitals.

All the parameters of the wave functions are optimized with the Stochastic reconfiguration method\cite{sor+01prb,sor+05prb} using the Correlated Sampling technique\cite{fil+00prb} to stabilize the convergence if necessary.
To better describe the  dynamical correlation among the particles, thus improving the binding energy estimates, we also employ the DMC\cite{fou+01rmp} method with the Fixed-Node approximation\cite{Reynolds1982}.
DMC is a projection method able to converge towards the exact ground state of a system through a time evolution.\cite{fou+01rmp}
In view of the sign-problem, which appears for fermionic systems, it is necessary to constrain the nodal surface of the wave function (the region where the wave function is zero) by introducing an error in the energy evaluation.\cite{fou+01rmp} 
Yet, it can be shown that FN-DMC, at least in the formalism employed here, is variational with respect to the quality of nodal surfaces, which here are defined by the optimized trial wave functions.   

\subsection{Computational details}
MP2 calculations for the positronic systems were carried out with the LOWDIN software \cite{Flores-Moreno2014} using the standard aug-cc-pVTZ basis set \cite{Dunning1989} for electrons with an uncontracted set of (6s4p3d2f) GTFs, denoted as PSX-TZ, for positrons \cite{moncada2020covalent}. VMC and DMC calculations were performed with the \texttt{QMeCha}\cite{QMeCha,cha+22jctc} QMC package, published privately on GitHub.
The electronic and positronic Slater determinants for the QMC calculations of \ce{H} systems were constructed from (7s3p2d) primitive GTFs contracted to [4s3p2d], with the initial values of the exponents and contraction coefficients taken from the aug-cc-pVTZ basis set \cite{Dunning1989,Kendall1992,Woon1993,Wilson1999}. 
In the VMC calculations, orbital coefficients, basis set contraction coefficients and GTF exponents were variationally optimized with the Stochastic Reconfiguration Optimization method \cite{Casula2003}, amounting to 1152 non-zero parameters for 2e$^+$[H$^{3-}_3$] in the singlet state, and 837 for 2e$^+$[H$^{3-}_3$] in the triplet state. 
The VMC optimized wavefunctions were taken as guiding functions in the fixed-node DMC calculations \cite{Reynolds1982} performed
with 6000 walkers divided into 5000 blocks, each 100 steps long, for a total of $3\times10^9$ sampled configurations.
Calculations were repeated with a time step of 0.005 a.u. to verify the accuracy of the DMC results. Statistical agreement was observed between the results computed with the two time steps. The DMC density plots were obtained by counting the number of particles of each weighted configuration in a three-dimensional grid. The final bonding properties such as energy minimum, equilibrium distances, and forces constants were estimated by fitting our VMC and DMC energy values to a 4th degree polynomial with respect to one of the internuclear distances. 

\section{Results and discussion}\label{results}
\subsection{Energy stability analysis}
\label{sectionCycles}

\begin{figure}[!ht]
    \centering
    \includegraphics[width=0.475\textwidth]{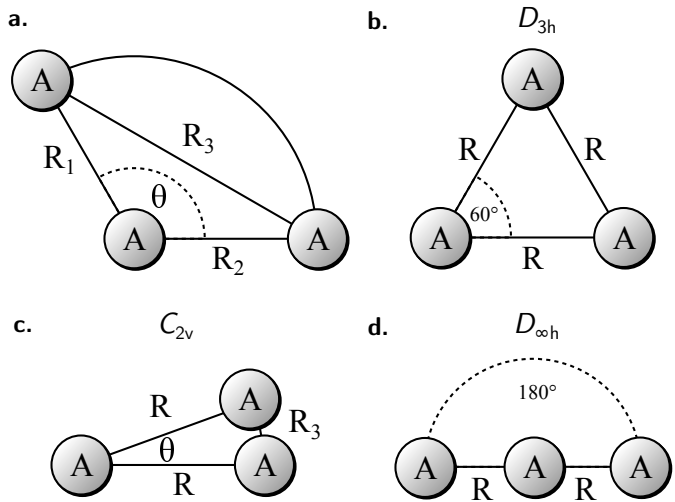}
    \caption{Panel \textbf{a} shows the coordinates for the molecules with three atomic centers. The other panels illustrate relevant particular cases: \textbf{b} equilateral triangle ($\ce{D$_{3h}$}$ symmetry), \textbf{c} isosceles triangle ($\ce{C$_{2v}$}$ symmetry), and \textbf{d} collinear geometry ($\ce{D_{\infty h}}$ symmetry). }
    \label{Fig:diagram}
\end{figure}

\begin{figure*}[hp!]
\includegraphics[width=.5\textwidth]{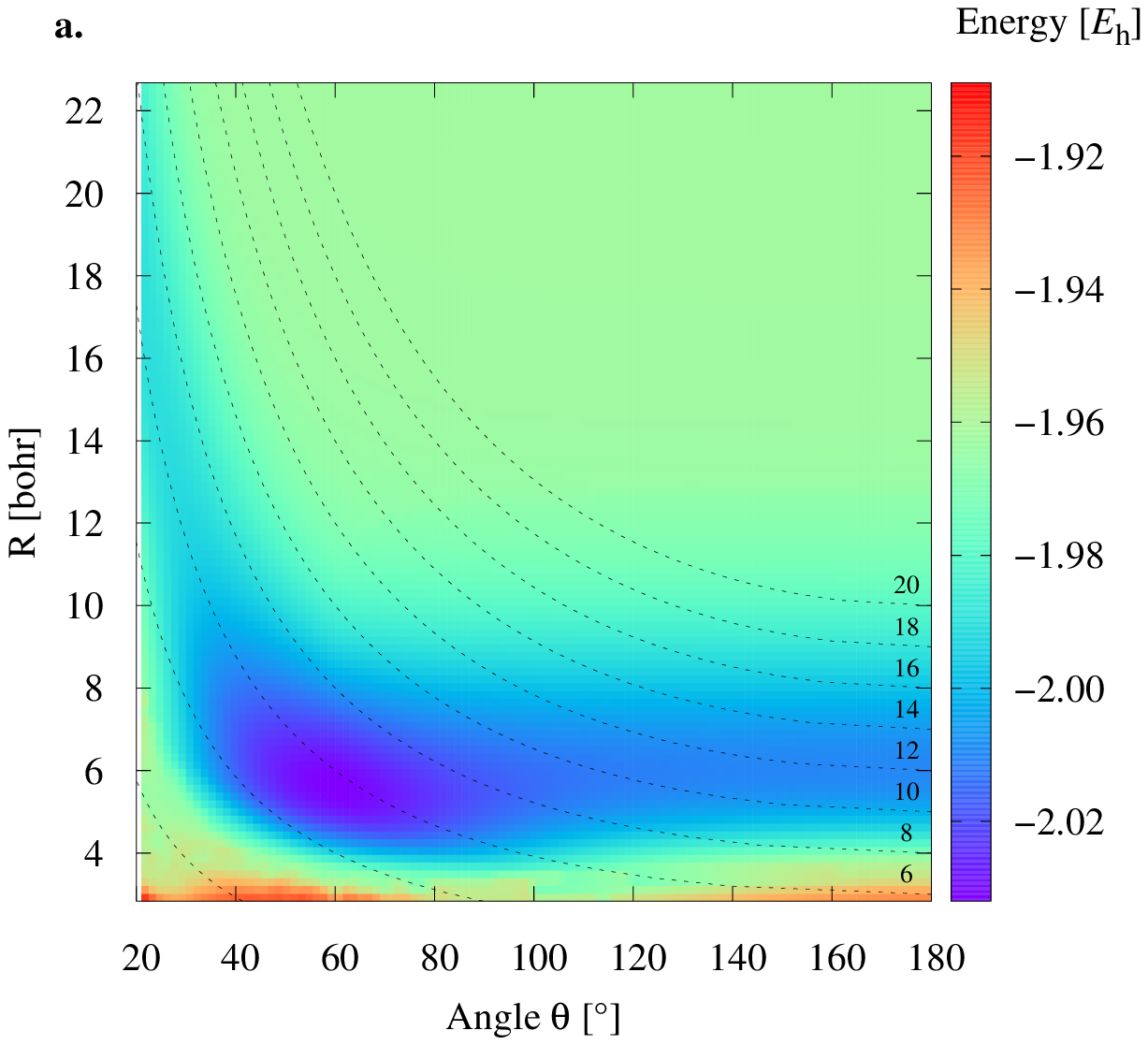}
\includegraphics[width=.5\textwidth]{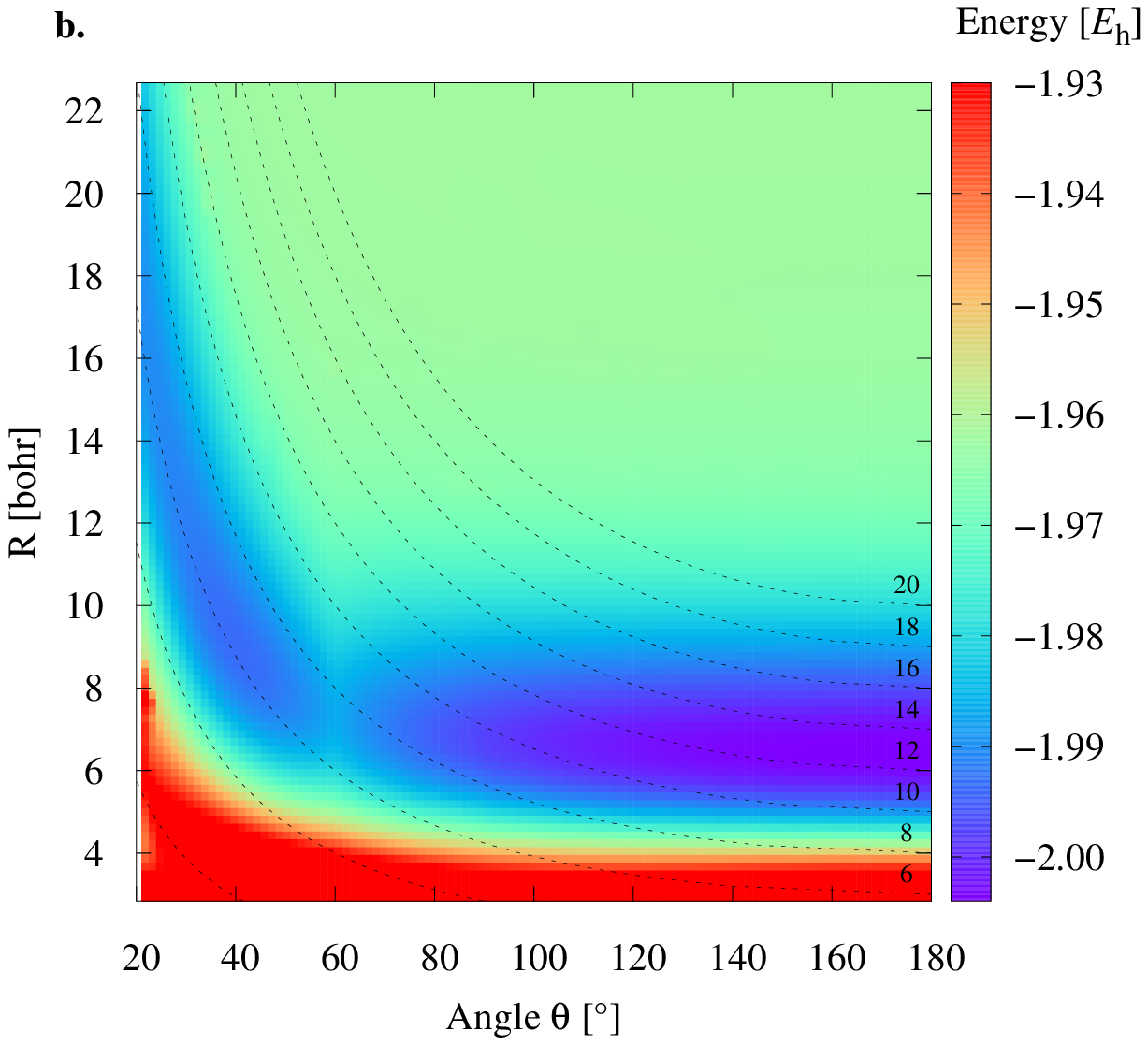}
\caption{Potential energy surfaces for the singlet (\textbf{a}) and triplet (\textbf{b}) positronic states of the 2\ce{e$^+$[H$^{3-}_3$]} system. The energies were computed at the MP2 level using the aug-cc-pVTZ/PSX-TZ combination of electronic and positronic basis sets centered at the hydrogen nuclei.
The dashed lines correspond to constant $R_3$ paths obtained for $\ce{C$_{2v}$}$ conformations.
}\label{Fig:2D_MP2_pos_singlet}
\end{figure*}

To study the 2\ce{e^+[H^{3-}_3]} system, we assume the electrons to be in a closed shell singlet state, while for the positrons we investigate both the singlet and triplet states.
The initial exploratory studies of the PES are performed at the MP2 level of theory along the coordinates $R_1=R_2=R$ and $\theta$, shown in Figure \ref{Fig:diagram}a, to preserve the symmetry of the charge distribution. 
The two-dimensional potential energy surfaces obtained for both the singlet and triplet positronic states in that system of coordinates (constrained to \ce{C$_{2v}$} symmetry) are shown in Figure \ref{Fig:2D_MP2_pos_singlet},
together with the contours for several $R_3$ constant values that are shown as dashed lines. 
The singlet state displays a single energy minimum with \ce{D$_{3h}$} symmetry (see Figure \ref{Fig:diagram}b), corresponding to a triangular configuration with $\theta=60^{\rm o}$ and $R\approx 6$ bohr. 
In contrast, the triplet state has a minimum at a linear conformation with D$_{\infty h}$ symmetry (see Figure \ref{Fig:diagram}d) and $R\approx 6$ bohr.

Further analysis of the PESs in Figure \ref{Fig:2D_MP2_pos_singlet} reveals that both the singlet and triplet states display two regions with low energies (blue color). The first region follows a dissociation path along the $R$ coordinate in the \ce{C$_{2v}$} configuration, with the $R_3$ coordinate constrained to 6 bohr. The second region corresponds to the transition between the \ce{D$_{3h}$} and \ce{D$_{\infty h}$} configurations along the angle coordinate, for constant $R$. The optimal MP2 geometries were further refined with the VMC and DMC methods for both the singlet and triplet states. The minima were located for geometries constrained to \ce{D$_{3h}$} and \ce{D$_{\infty h} $} symmetries. The results shown in Table \ref{tab:total_energies} corroborate the singlet \ce{D$_{3h}$} state as the most stable one. Below, we focus the discussion on the lowest-energy 2\ce{e^+[H^{3-}_3]} singlet state. 


\begin{table}[]
 \centering
\caption{Equilibrium energies and distances computed with the VMC and DMC methods for the singlet and triplet states of the 2\ce{e$^+$[H$^{3-}_3$]} system. The symmetry of the potential energy minima is also indicated.}
\begin{tabular}{cccll}
\hline
{\begin{tabular}[c]{@{}c@{}}Positronic \\ State\end{tabular}} & \multicolumn{1}{c}{{Symmetry}} & {Method} & \multicolumn{1}{c}{{\begin{tabular}[c]{@{}c@{}}Energy\\ $[E_{\text{h}}]$\end{tabular}}} & \multicolumn{1}{c}{{\begin{tabular}[c]{@{}c@{}}R$_{eq}$\\ $[$bohr$]$\end{tabular}}} \\ \hline
\multirow{2}{*}{Singlet}  & \multirow{2}{*}{\ce{D$_{3h}$}}         & VMC             & $-$2.1489(1) & 6.15(1) \\
                          &                                        & DMC             & $-$2.1652(2) & 6.11(1) \\
\multirow{2}{*}{Triplet}  & \multirow{2}{*}{\ce{D$_{\infty h} $}}  & VMC             & $-$2.1266(1) & 6.77(2) \\
                          &                                        & DMC             & $-$2.1401(4) & 6.62(1) \\ 
                          \hline
\end{tabular}
\label{tab:total_energies}
\end{table}

We explored the thermodynamic stability of 2\ce{e^+[H^{3-}_3]} with respect to the dissociation channels given below, 
\begin{align}
\ce{2e+[H^{3-}_3]}\longrightarrow
 \begin{cases}
a)\ \ce{2e+[H^{2-}_2]} +\ce{H^-} & \Delta E_{a}=48.6(2) \ce{m$E$_h}\\
b)\ \ce{e+[H^{2-}_2]} +\ce{PsH} & \Delta E_{b}=35.5(2) \ce{m$E$_h}\\
c)\ \ce{H_2} +\ce{PsH}+\ce{Ps^-} & \Delta E_{c}=-60.5(2) \ce{m$E$_h} \\
d)\ \ce{H^{-}_3} +\ce{Ps2} & \Delta E_{d}=-54.4(2) \ce{m$E$_h}. \\
\end{cases}
\label{adiabatic}
\end{align}
The dissociation energies of channels \ref{adiabatic}$a$-\ref{adiabatic}$d$ were calculated employing the 
DMC results of Table \ref{tab:total_energies} and the energy data reported in Table S1.
The negative $\Delta E$ of channels \ref{adiabatic}$c$ and \ref{adiabatic}$d$ reveal that 2\ce{e^+[H^{3-}_3]} is thermodynamically unstable with respect to these dissociation products.
It is therefore clear that the lowest energy arrangement of a set of three protons, six electrons, and two positrons consists of a sum of dissociated species rather than the bonded \ce{D$_{3h}$} structure.

A similar situation was found for positron bonded dihydrides. Previous studies pointed out that I) \ce{e+[H^{2-}_2]}\cite{Bressanini2021,Ito2020,cha+22jctc}, II) 2\ce{e+[H^{2-}_2]} \cite{Bressanini2021-2}, and III) 3\ce{e+[H^{2-}_2]}\cite{3e+} are thermodynamically unstable with respect to dissociation into I') \ce{H_2 + Ps^-}, II') \ce{H_2 + Ps_2} and III') \ce{H_2 + Ps_2 + e^+}, respectively. The local range of stability for the systems I, II and III was explored by comparing the potential energy curves (PECs) for the (I, I'), (II, II') and (III,III') pairs. The PEC pairs were found to intersect at internuclear distances considerably shorter than the equilibrium distances of the corresponding positron bonded dihydrides. Those studies further confirmed the kinetic (or local) energy stability of the  dihydrides I-III, since their energy barriers, calculated at the crossing points of the potential curves, were found to be sufficiently high to support a few vibrational states. 

Along similar lines, we further explore the local stability of 2\ce{e^+[H^{3-}_3]} around the \ce{D$_{3h}$} minimum. 
We consider the four lowest-energy vertical detachment channels given by
\begin{align}
2\ce{e+[H^{3-}_3]}\longrightarrow
 \begin{cases}
 a)\ \ce{2e^+[H^{2-}_3]} + \ce{e^-}   & \Delta E_{a} = 72.8(2) \ce{m$E$_h}\\
 b)\ \ce{e^+[H^{2-}_3]} + \ce{Ps}   &  \Delta E_{b} = 68.1(2) \ce{m$E$_h}\\ 
 c)\ \ce{e^+[H^-_3]} +  \ce{Ps^-} &  \Delta E_{c} = 116.8(2) \ce{m$E$_h}\\ 
 d)\ \ce{H^-_3} + \ce{Ps_2}  & \Delta E_{d} = 107.4(2) \ce{m$E$_h}.\\
 \end{cases}\label{vertical}
 \end{align}
Here, positive $\Delta E$ values confirm the stability of 
2\ce{e^+[H^{3-}_3]} against vertical detachments.

\begin{figure*}[hp!]
\includegraphics[width=\textwidth]{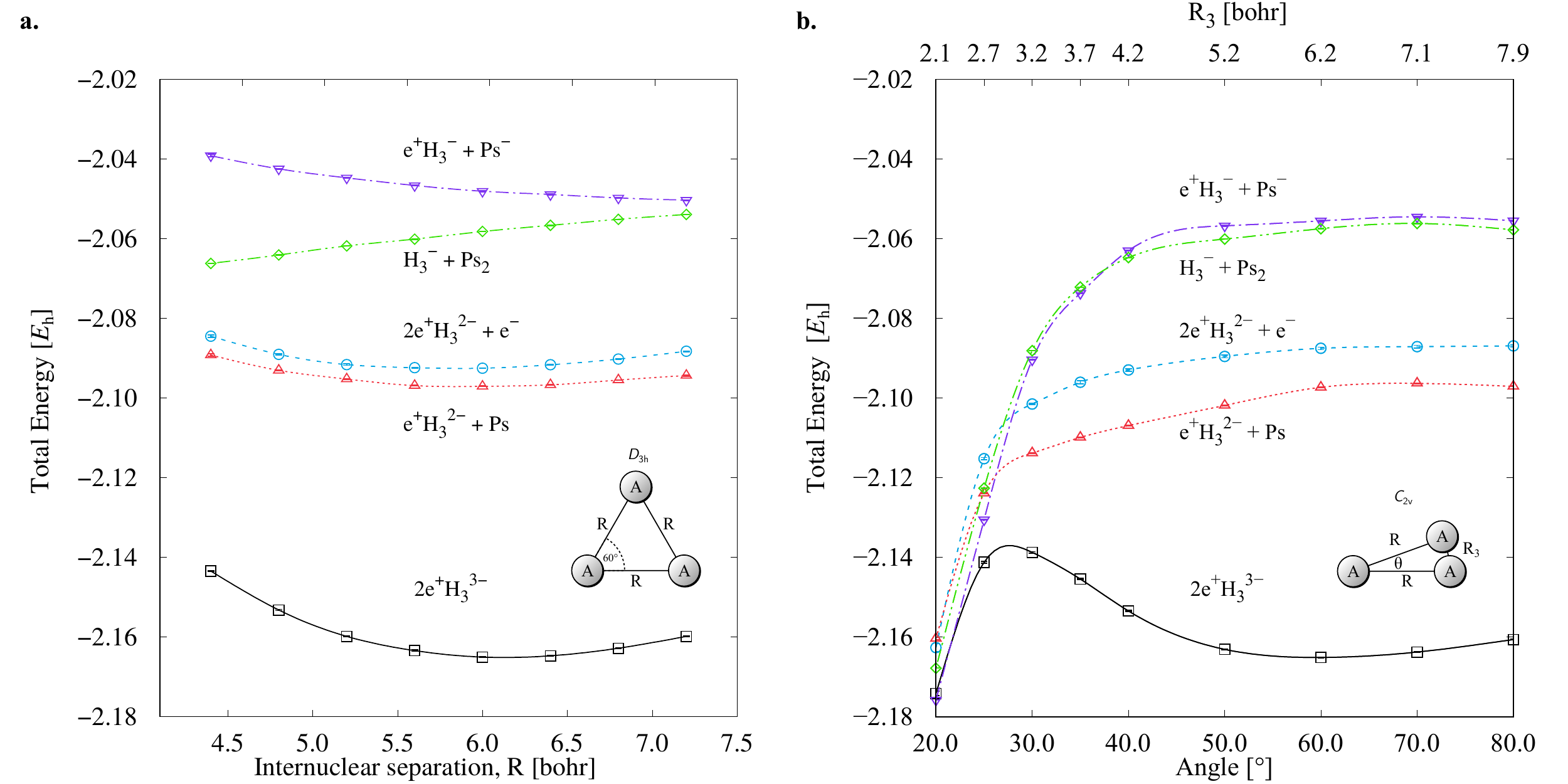}
\caption{Potential energy curves for the 2\ce{e^+[H^{3-}_3]} system and the vertical detachment channels \ref{vertical}$a$-\ref{vertical}$d$. The energies were computed with the DMC method as functions of the $R$ coordinate in the \ce{D_${3h}$} symmetry (\textbf{a}), and of the \ce{\theta} coordinate in the \ce{C_{2v}} symmetry (\textbf{b}) fixing R=6.1 bohr. We employed the lowest energies obtained from variational calculations for \ce{Ps^-} and \ce{Ps_2} \cite{Korobov2000,Bubin2006}, and the exact value for Ps (-0.25 \ce{E_h}).}
\label{Fig:dissociations}
\end{figure*}

Figure \ref{Fig:dissociations} presents PECs along the $R$ coordinate constraining the system to preserve the \ce{D_${3h}$} symmetry, 
and along the $\theta$ coordinate imposing \ce{C_{2v}} symmetry with $R=6.1$ bohr.
Panel \ref{Fig:dissociations}a reveals that the energy of 2\ce{e^+[H^{3-}_3]} along the $R$ coordinate is always lower than those computed for the \ref{vertical}$a$-\ref{vertical}$d$ channels, and no curve crossings were found.
In contrast, panel \ref{Fig:dissociations}b reveals that the PEC of 2\ce{e^+[H^{3-}_3]} intersects the \ref{vertical}$c$ potential around 20$^{\rm o}$ and $R_3=2.1$ bohr. 
The stabilization of the \ref{vertical}$c$ channel as the $R_3$ distance decreases is consistent with the formation of a H$_2$ molecule ($R=1.40$ bohr) in the adiabatic channel \ref{adiabatic}$c$.
Although the \ref{adiabatic}$c$ channel energy lies below the 2\ce{e^+[H^{3-}_3]} energy, the intersections are separated from the \ce{D_${3h}$} minimum by a rather high ($\approx 20$ m$E_{\text{h}}$) and broad ($\theta\approx 30^{\rm o}$) energy barrier across the 2\ce{e^+[H^{3-}_3]} potential. 
The PECs in Figures \ref{Fig:dissociations}a and \ref{Fig:dissociations}b therefore corroborate the local (kinetic) energy stability of the 2\ce{e^+[H^{3-}_3]} species.
 
\subsection{Comparison with analogous purely electronic systems}
As in previous studies of positron bonded dihydrides\cite{charry2018binding} and dihalides\cite{moncada2020covalent}, we compare the properties of the 2\ce{e^+[H^{3-}_3]} system with 
two 3c2e bonded systems, i.e., \ce{H^+_3} and \ce{Li^+_3}. 
Specifically, the equilibrium distances and the force constants for the symmetric stretching ($\nu_1$) and bending ($\nu_2$) vibrational modes of the \ce{D_${3h}$} ground states, calculated with the DMC method, are shown in Table \ref{tab:opt_geom}. 
The equilibrium distance of the positronic molecule ($R=6.11$ bohr) is much closer to that of \ce{Li^+_3} ($R=5.64$ bohr) that \ce{H^+_3} ($R=1.63$ bohr). 
This result is consistent with our earlier studies, which pointed out that purely electronic molecules with isoelectronic cores are the closest analogues to positron bonded systems. Although larger discrepancies are found for the vibrational frequencies of 2\ce{e^+[H^{3-}_3]} and \ce{Li^+_3} (roughly a factor of 2), they are still much closer if compared to the frequencies of \ce{H^+_3}, having different orders of magnitude. The present calculations for three-center two-particle bonds therefore corroborate the previous findings for diatomic systems.

\begin{table}[ht!]
 \centering
\setlength\extrarowheight{3pt}
\caption{Total energies (in \ce{$E$_h}), equilibrium distance (in bohr) and force constants (in a.u.) for the symmetric stretching $\nu_1$ mode and bending $\nu_2$ mode of triatomic systems in \ce{D$_{3h}$} symmetry calculated at DMC level. }
\begin{tabular}{lllll}
\hline
{System}        & \multicolumn{1}{c}{{$E$}} & \multicolumn{1}{c}{{$R$}} & \multicolumn{1}{c}{\textbf{$k_{\nu_1}$}} & \multicolumn{1}{c}{\textbf{$k_{\nu_2}$}} \\ \hline
\ce{H^{+}_3} & $-$1.346(4) & 1.63(3)  & 0.57(5)  & 0.16(2)             \\
\ce{Li^{+}_3} & $-$22.3419(1) & 5.645(7) & 0.0254(3) & 0.0110(4)     \\
2\ce{e^+[H^{3-}_3]}  & $-$2.1652(2) & 6.11(1) & 0.0114(4)    & 0.0047(7)      \\
\hline
\end{tabular}
\label{tab:opt_geom}
\end{table}

The longer equilibrium distances and lower vibrational frequencies also indicate a weaker 3c2p bond compared to the 3c2e analogues. 
In addition, we contrast the energy stability of 2\ce{e^+[H^{3-}_3]} with those of the purely electronic \ce{Li^+_3}, and \ce{H^+_3}. To that aim, we consider the lowest energy chemically meaningful dissociation channels of \ce{Li^+_3}, and \ce{H^+_3} 

\begin{align}
\ce{H^+_3}\longrightarrow
 \begin{cases}
a)\ \ce{H2 + H+}&E_{diss}=171(4) \ \ce{m$E$_h}\\
b)\  \ce{H^+2 + H}&E_{diss}=243(4)\ \ce{m$E$_h}
\label{channel1}
 \end{cases}\\
\ce{Li^+_3}\longrightarrow
 \begin{cases}
a)\ \ce{Li2 + Li+}&E_{diss}=66.8(1)\ \ce{m$E$_h}\\
b)\ \ce{Li^+2 + Li}&E_{diss}=58.3(1)\ \ce{m$E$_h}
  \label{channel2}
 \end{cases}\\
  2\ce{e^+[H^{3-}_3]}\longrightarrow
 \begin{cases}
a)\ \ce{2e^+ [H^{2-}_2] + [H^-]}  &E_{diss}=48.6(2)\ \ce{m$E$_h}\\
b)\ \ce{e^+ [H^{2-}_2] + e^+[H^-]} &E_{diss}=35.5(2)\ \ce{m$E$_h}
  \label{channel3}
 \end{cases}
 \end{align}
The energies of channels \ref{channel1} and \ref{channel2} 
were calculated employing the DMC energy data reported in Table S1.
We observe that the lowest energy dissociation channel for \ce{H^+_3} ($E_{diss}=171.84 $m$E_h$) involves the formation of a neutral molecule and a proton. Contrastingly, for \ce{Li^+_3} ($E_{diss}=58.27 $m$E_h$) and 2\ce{e^+[H^{3-}_3]} ($E_{diss}=35.54 $m$E_h$), involves the formation of a neutral atom, and an ionic single particle bonded molecule. It is worth noticing that $E_{diss}$ of \ce{Li^+_3} and 2\ce{e^+[H^{3-}_3]} are of the same order and differ considerably from that of \ce{H^+_3} for all dissociation channels considered here. 

\subsection{Densities}\label{sectionDens}
\begin{figure}[!ht]
\includegraphics[width=.5\textwidth]{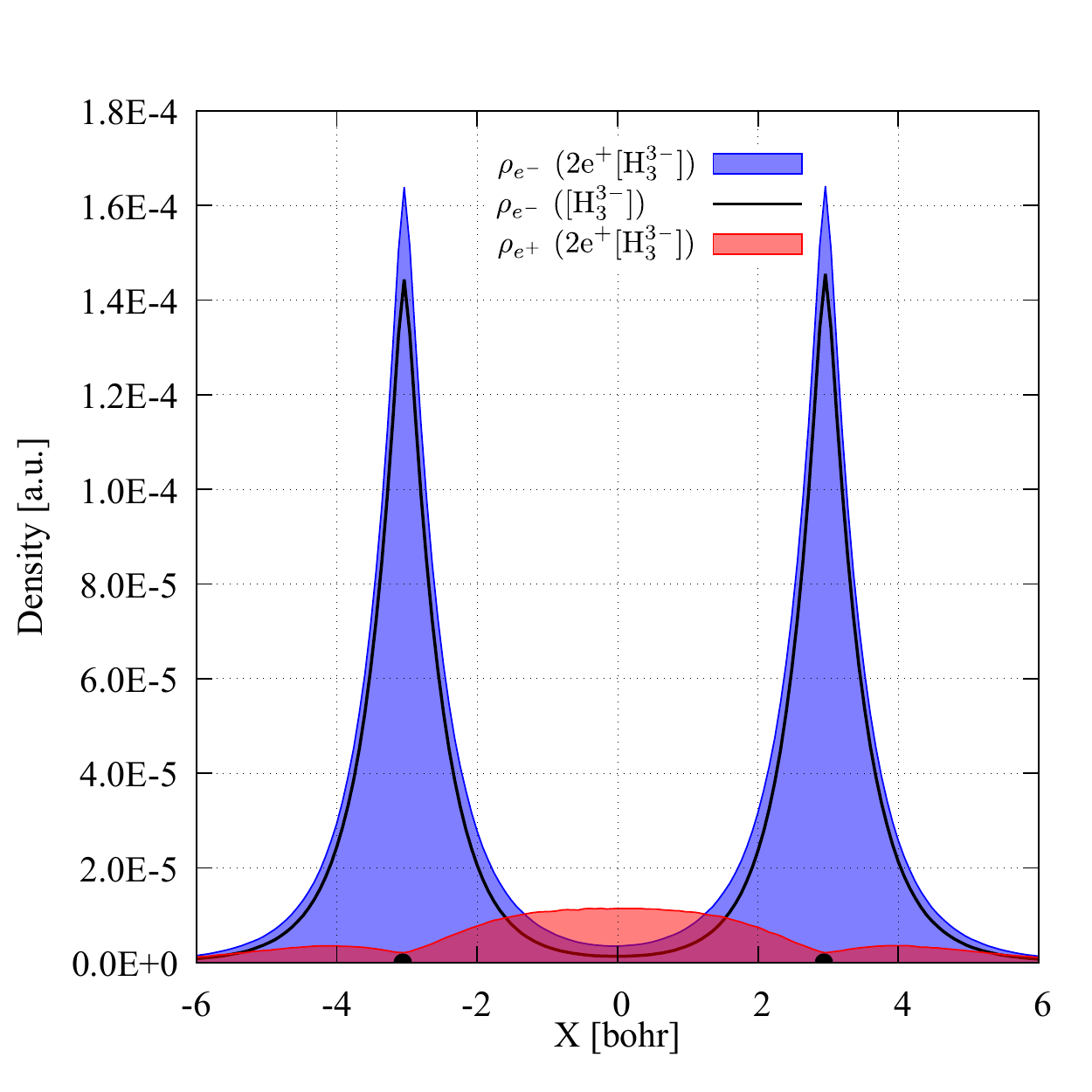}
\caption{
1D cuts of the electronic densities of the \mbox{[H$^{3-}_3$]} (black line) and \mbox{2e$^+$[H$^{3-}_3$]} (blue) systems. The 1D positron density of the latter compound is also shown (red). 
The densities were obtained as histograms of the number of particles inside voxels (width=0.06 bohr) disposed along an internuclear axis. 
Electronic and positronic densities are normalized to 6 and 2, respectively.
}
\label{Fig:1d_densities}
\end{figure}

To gain further insight into the 3c2p bond formation, we computed the positron ($\rho_{e^+}$) and electron ($\rho_{e^-}$) densities of 2\ce{e^+[H^{3-}_3]}. One-dimensional (1D) cuts of the electron densities obtained for the unbound \ce{[H^{3-}_3]} trianion and the bound 2\ce{e^+[H^{3-}_3]} compound are shown in Figure \ref{Fig:1d_densities}. The densities are remarkably similar, thus pointing out that the binding mechanism in 2\ce{e^+[H^{3-}_3]} is not electronic, as previously observed for positronic covalent bonds \cite{charry2018binding,Goli2019,moncada2020covalent}. In contrast, the positron density prominently accumulates between the H nuclei, balancing the otherwise repulsive interaction between the anions.

\begin{figure*}[!ht]
\includegraphics[width=.33\textwidth]{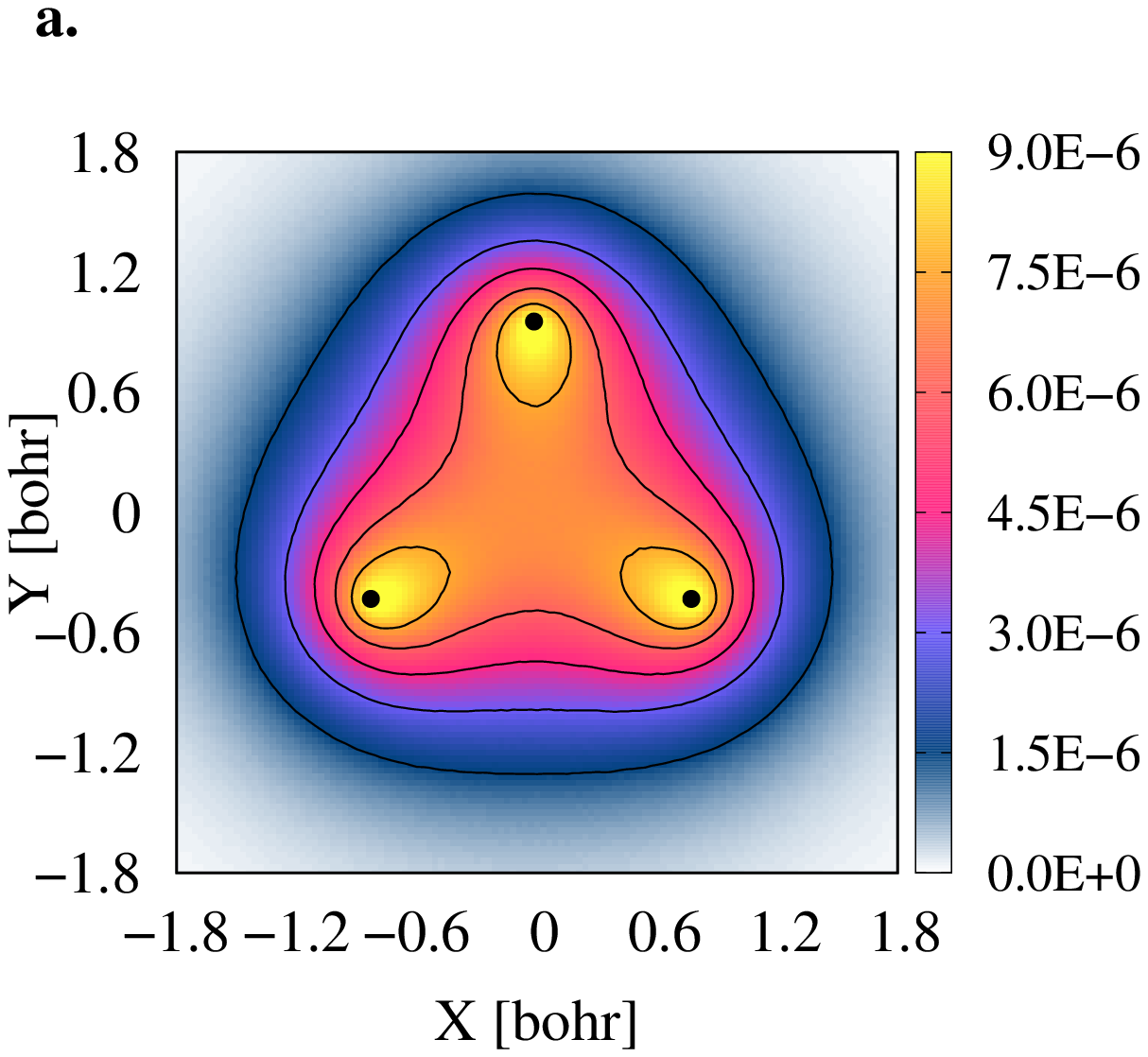}
\includegraphics[width=.33\textwidth]{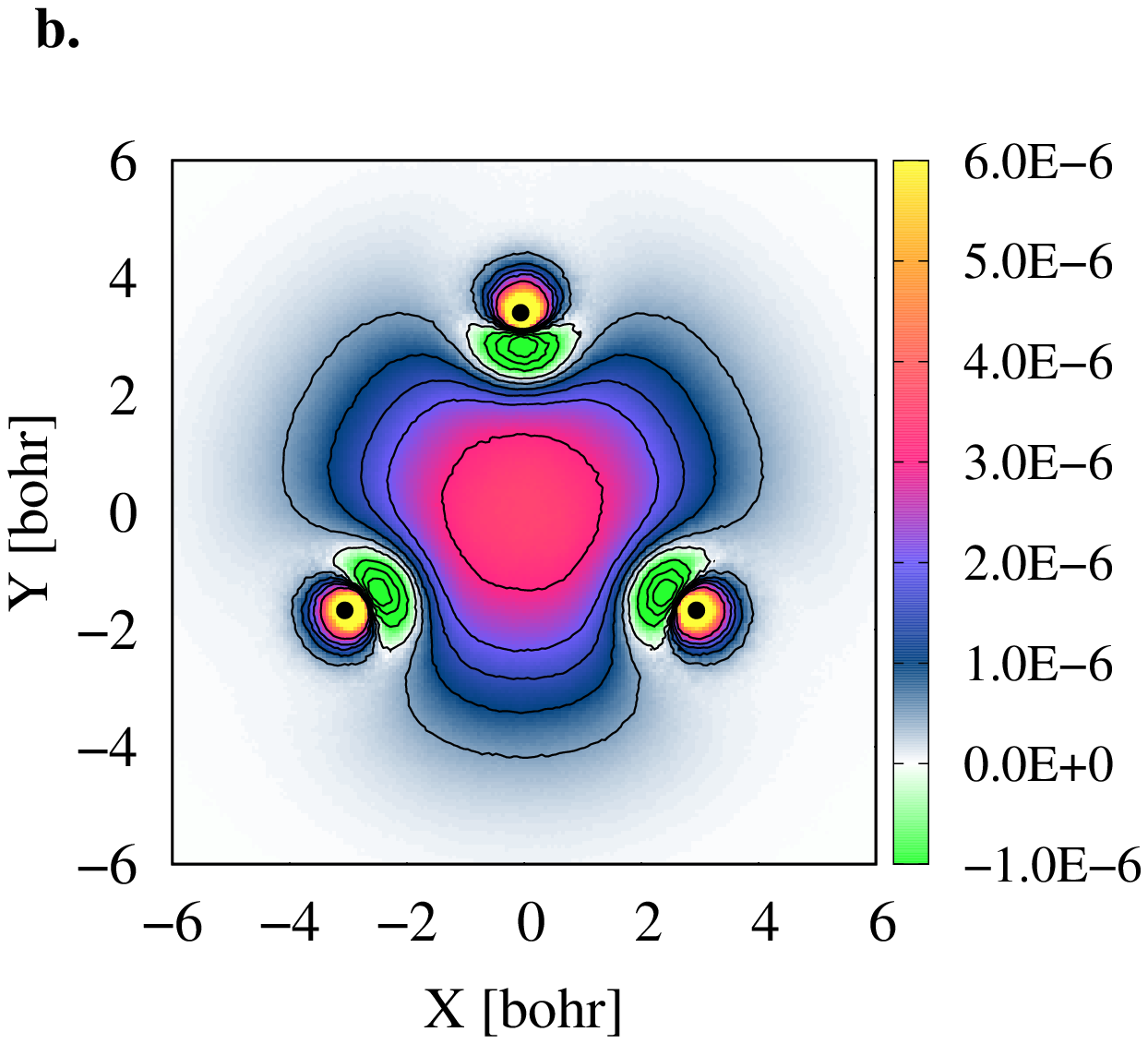}
\includegraphics[width=.33\textwidth]{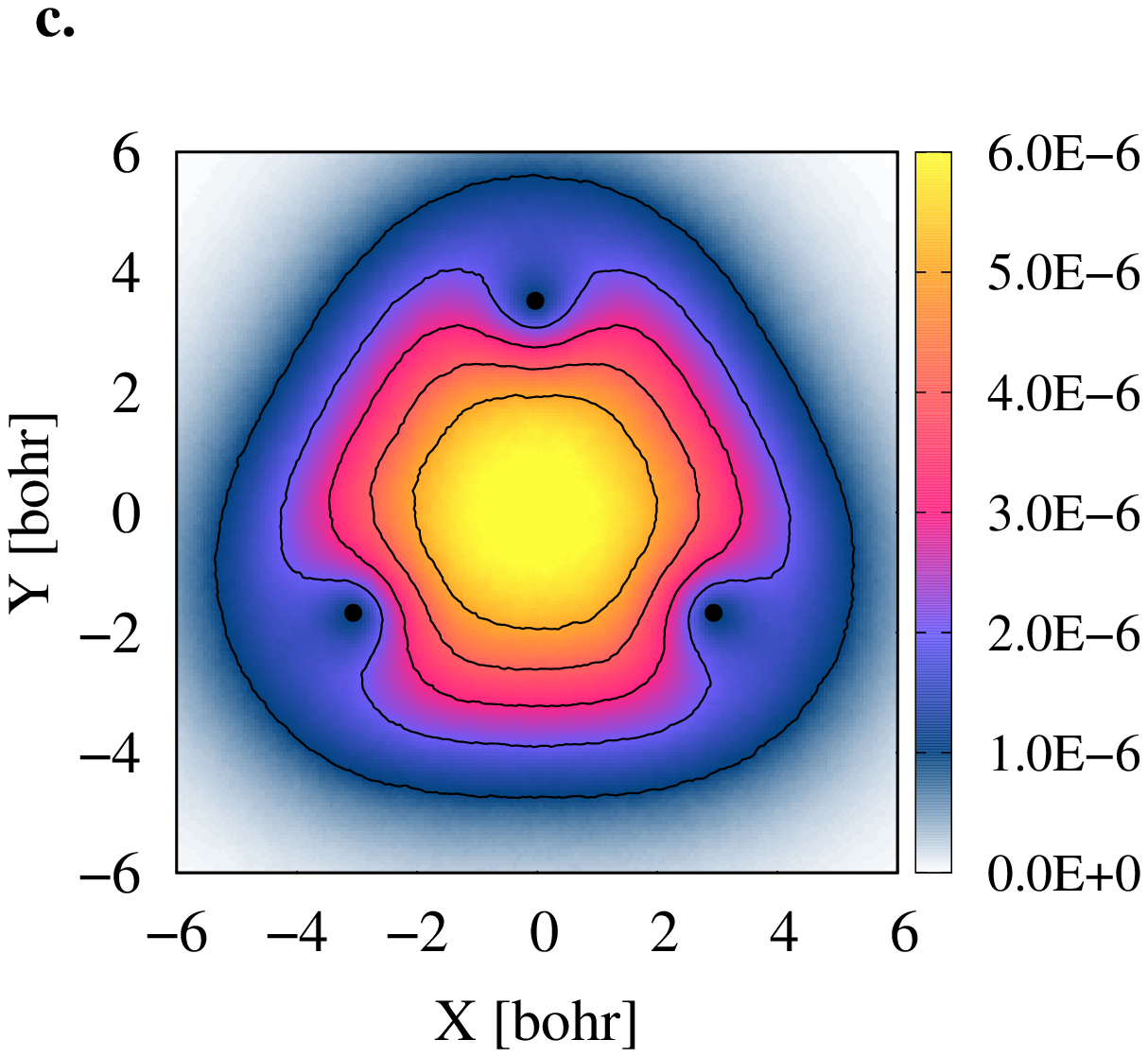}
\caption{Comparison of three-center two-particle bonding densities. Electronic density of \mbox{2e$^-$[H$^{3+}_3$]} (\textbf{a}), electronic bonding density of \mbox{[Li$^{3+}_3$]} (\textbf{b}), and positronic bonding density of \mbox{2e$^+$[H$^{3-}_3$]} (\textbf{c}). 
The densities were obtained with the DMC method as histograms of the number of particles inside voxels (width=0.06 bohr) disposed along the molecular plane.
In all panels, the atomic nuclei are represented as black dots.}

%
%
\label{Fig:2ddensities}
\end{figure*}

A comparison between the two-particle bonding densities in three-center systems is carried out in Figure \ref{Fig:2ddensities}. 
In \ce{H^{+}_3} the bonding density presented in panel a) is the total electron density, $\rho_{e^-}$. $\rho_{e^-}$ in \ce{H^{+}_3} has maxima on top of the nuclei and accumulates around the centroid of the system, where $\rho_{e^-}$ has a local minimum.
For \ce{Li^{+}_3} the bonding density, $\Delta \rho_{e^-}$, displayed in panel b), was obtained as the $\rho_{e^-}$ difference between the \ce{[Li^{+}_3]} and \ce{[Li^{3+}_3]} systems, both calculated at the same geometry.
$\Delta \rho_{e^-}$ in \ce{Li^{+}_3} also displays maxima around the nuclei and accumulates around the centroid, but in this case, has a local maximum at that point.
The symmetric accumulation of electronic density in the centroid of the system is a signature of  3c2e bonding.
In turn, in 2\ce{e^+[H^{3-}_3]} the bonding density is the total positron density, $\rho_{e^+}$ portrayed in panel c).
$\rho_{e^+}$ in 2\ce{e^+[H^{3-}_3]} is depleted around the nuclei, as expected from the Coulomb interaction, gradually accumulates at the internuclear region and peaks at the centroid, clearly resembling $\Delta \rho_{e^-}$ of \ce{Li^{+}_3} in that region.
Therefore, this accumulation around the centroid of 2\ce{e^+[H^{3-}_3]} can be viewed as the formation of a 3c2p bond.

The densities (Figure \ref{Fig:2ddensities}), vibrational properties (Table \ref{tab:opt_geom}) and bond energies (Eqs. \ref{adiabatic} and \ref{channel2}) provide strong evidence that similar three-center two-particle bonding mechanisms are present in \ce{Li^{+}_3} and in 2\ce{e^+[H^{3-}_3]}.

Now that we have established the local stability of 2\ce{e^+[H^{3-}_3]}, its bonding properties and its similarities with \ce{Li^{+}_3}, it is worth exploring the chemistry of the latter to gain further insight into the chemical implications of the three-center two-positron bond. 

Previous studies of the nature of the 3c2e bond in \ce{Li^{+}_3} based on
topological analyses of the electronic density \cite{Havenith2005,Foroutan-Nejad2009} and an Interference Energy analysis\cite{DeSousa2022} have concluded that \ce{Li^{+}_3} can be considered as the smallest metallic cluster, where the valence electrons in the centroid act as free metallic electrons.
Therefore, the reported similarities of the positronic bonding density of 2\ce{e^+[H^{3-}_3]} and the 3c2e bond in \ce{Li^{+}_3}, suggest that the positrons in 2\ce{e^+[H^{3-}_3]} could also be considered as `pseudo'-metallic in character. 

Additionally, to explain the relative stability and structure of \ce{Li^{+}_3}, the presence of $\sigma$-aromatic or $\sigma$-antiaromatic states is still widely discussed  \cite{Alexandrova2003,Havenith2005,DeSousa2022,He2022}. 
A similar discussion could be extended to interpret the positronic delocalization in 2\ce{e^+[H^{3-}_3]} as evidence of a new type of positronic $\sigma$-aromaticity or $\sigma$-antiaromaticity. 

The extension of the metallic bond and aromatic states concepts to positronic molecules requires a more in-depth analysis of the bonding nature, which could be performed with either energy decomposition schemes, interference analysis \cite{DeSousa2022}, or through the investigation of its topological properties (MC-QTAIM) \cite{Goli2011,shant2017,Goli2019}. However, those studies are beyond the scope of the present work. 

\section{Conclusions}\label{conclusions}

Theoretical studies have evidenced the formation of locally stable systems composed of two repelling atomic anions and one-to-three positrons. 
The anti-particles are mainly responsible for the conformers' stabilization, thus expanding the definition of chemical bonding beyond the ordinary purely electronic systems\cite{charry2018binding,Goli2019,Ito2020,cha+22jctc,Bressanini2021,Bressanini2021-2,3e+}. 
In this work, we have found that two positrons can form a locally stable bound state with three hydride anions in a D$_{3h}$ configuration and singlet spin state. 
The results suggest the formation of a 3c2p bond, pointing out that the chemical bond concept reaches beyond the positron covalent bonds in two-center molecules, addressed in previous studies. 
The dissociation products $\ce{H_2} +\ce{PsH}+\ce{Ps^-}$ correspond to the lowest-energy configuration of the system comprising two positrons, six electrons and three hydrogen nuclei. Nevertheless, our vertical detachment and adiabatic energy dissociation analysis confirm the local stability of 2\ce{e^+[H^{3-}_3]} around the D$_{3h}$ equilibrium geometry. 
A comparative analysis reveals that the positronic molecule is similar to the trilithium cation with isoelectronic atomic cores, although it remarkably differs from the trihydrogen cation. 
Similarities were found for the equilibrium geometries, force constants for symmetric stretch and bending modes, and also dissociation energies trends. 
In addition, the electronic (\ce{Li_3^+}) and positronic (2\ce{e^+[H^{3-}_3]}) bonding densities are similar around the internuclear region and equally spread among all three atomic centers, suggesting strong similarities between the 3c2p and 3c2e bonds. The positronic species has longer bond lengths, weaker force constants and lower bond energies than the trilithium cation, which can be viewed as the analogue purely electronic system. This trend does not seem to be general, as it was found for positronic hydrides but not for positronic dihalides.

The present results extend the findings of previous studies to a new type of positron bond between otherwise repelling atomic anions. 
Since positron bonded systems are fundamentally distinct from those formed by positron attachment to stable atoms and molecules, these results will hopefully stimulate further theoretical and experimental researches on novel physical and chemical processes involving those antiparticles.

\section*{Author Contributions}

JC: Conceptualization, Formal Analysis, Investigation, Methodology, Resources, Software, Visualization, Writing - original draft, Writing - review \& editing.
FM: Conceptualization, Formal Analysis, Investigation, Methodology, Resources, Software, Visualization, Writing - original draft, Writing - review \& editing
MB: Formal Analysis, Methodology, Resources, Software, Writing - original draft, Writing - review \& editing
LPG: Conceptualization, Investigation, Methodology, Writing - original draft, Writing - review \& editing
MTNV: Funding acquisition, Methodology, Validation, Writing - original draft, Writing - review \& editing
AT: Funding acquisition, Methodology, Resources, Software, Validation, Writing - original draft, Writing - review \& editing
AR: Conceptualization, Funding acquisition, Investigation, Methodology, Project administration, Resources, Software, Supervision, Validation, Writing - original draft, Writing - review \& editing

\section*{Conflicts of interest}
There are no conflicts to declare.

\section*{Acknowledgements}
FM acknowledges financial support from the Colombian Ministry of Sciences (Minciencias) through the "Doctorados Nacionales (757)" program.
MTNV acknowledges support from S\~ao Paulo Research Foundation, FAPESP (grant no. 2020/16155-7), and National Council for Scientific and Technological Development, CNPq (grant no. 304571/2018-0).
JACM acknowledges financial support from the Luxembourg National Research Fund (AFR PhD/19/MS, 13590856). MB acknowledges financial support from the Luxembourg National Research Fund (INTER/DFG/18/12944860). 
Part of the calculations presented in this paper were carried out using the HPC facilities of the University of Luxembourg~\cite{VBCG_HPCS14} {\small (see \href{http://hpc.uni.lu}{hpc.uni.lu})}.



\balance


\bibliography{bibliography} 
\bibliographystyle{rsc} 

%
%
%
%
\newcommand*\mycommand[1]{\texttt{\emph{#1}}}

%
%
\renewcommand{\thefigure}{\textbf{S\arabic{figure}}}
\renewcommand*{\tablename}{\bf{Table}}
\renewcommand{\thetable}{\textbf{S\arabic{table}}}

\setcounter{figure}{0}    
\setcounter{table}{0}    
%


\title{The three-center two-positron bond \\
Supplementary Information}
\author[1]{Jorge Charry}
\author[2]{F\'elix Moncada}
\author[3]{Matteo Barborini}
\author[4]{Laura Pedraza-Gonz{\'a}lez}
\author[5]{M\'arcio T. do N. Varella}
\author[6]{Alexandre Tkatchenko}
\author[7]{Andres Reyes}

\date{}
\affil[1,3,6]{\textit{Department of Physics and Materials Science, University of Luxembourg, L-1511 Luxembourg City, Luxembourg.}}
\affil[2]{\textit{Department of Physics, AlbaNova University Center, Stockholm University, S-106 91 Stockholm, Sweden.}}
\affil[4]{\textit{Department of Chemistry and Industrial Chemistry, University of Pisa, Via Moruzzi 13, 56124, Pisa, Italy. }}
\affil[5]{\textit{Instituto de Física, Universidade de São Paulo, Rua do Matão 1731, São Paulo, São Paulo, 05508-090, Brazil. }}
\affil[7]{\textit{Department of Chemistry, Universidad Nacional de Colombia, Av. Cra 30\#45–03, Bogot\'a, Colombia; E-mail: areyesv@unal.edu.co}}

\maketitle

\begin{table*}
\centering
\caption{Total energies (in \ce{E_h}) of atomic, diatomic and triatomic species along with diatomic and triatomic  equilibrium distances (in bohrs)}			
\label{tab:diss_species}
\begin{tabular}{lllll} \hline										
System	&	E(DMC)	&		&	E(Ref.)		&		\\ \hline
\ce{Ps}	&	-	&		&	-0.25$^a$		&		\\
\ce{Ps^-}	&	-	&		&	-0.262005	\cite{Korobov2000}	&		\\
\ce{Ps_2}	&	-	&		&	-0.516004	\cite{Bubin2006}	&		\\
\ce{H^-}	&	-0.52759(4)	&		&	-0.52779(3)	\cite{Ito2020}	&		\\
\ce{PsH}	&	-0.78919(4)	&		&	-0.78907(7)	\cite{Ito2020}	&		\\
\ce{Li^+}	&	-7.27992(1)	&		&	-7.279910(5)	\cite{Maldonado2010}	&		\\
\ce{Li}	&	-7.47801(2)	&		&	-7.47802(1)	\cite{Brown2007}	&		\\ \hline
System	&	E(DMC)	&	R(DMC)	&	E(Ref.)		&	R(Ref)	\\ \hline
\ce{H^+_2}	&	-	&		&	-0.602635	\cite{Guan2003}	&	2.00	\\
\ce{H_2}	&	-	&		&	-1.174476	\cite{CENCEK2008}	&	1.40	\\
\ce{H^+_3}$^b$	&	-1.346(4)	&	1.63(3)	&	-1.343426	\cite{Turbiner2013}	&	1.65	\\
\ce{H^-_3}$^c$	&	-	&		&	-1.703511	\cite{Ayouz2010}	&	1.42, 6.07	\\
\ce{Li^{+}_2}	&	-14.80562(2)	&	5.90	&	-14.80562(1)	\cite{Nasiri2017}	&	5.877	\\
\ce{Li_2}	&	-14.99175(6)	&	5.05	&	-14.9952(1)	\cite{Bressanini2005}	&	5.051	\\
\ce{Li^{+}_3} $^b$	&	-22.3419(1)	&	5.645(7)	&	-		&		\\
\ce{Li^{+}_3} $^d$	&	-22.31280(7)	&	5.937(7)	&	-		&		\\
\ce{e^+[H^{2-}_2]}	&	-1.3403(1)	&	6.39(3)	&	-1.3403(1)	\cite{Bressanini2021}	&	6.4(4)	\\
2\ce{e^+[H^{2-}_2]}	&	-1.5885(1)	&	6.003(7)	&	-1.5888(1)	\cite{Bressanini2021-2}	&	6.0(4)	\\
2\ce{e^+[H^{3-}_3]} $^b$	&	-2.1652(2) 	&	6.11(1) 	&			&		\\
2\ce{e^+[H^{3-}_3]} $^d$	&	-2.1401(4)	&	6.62(1)	&			&		\\ \hline
\multicolumn{5}{l}{$^a$ Exact} \\										
\multicolumn{5}{l}{$^b$ Singlet D$_{3h}$ symmetry} \\										
\multicolumn{5}{l}{$^c$ Singlet C$_{\infty v}$ symmetry} \\										
\multicolumn{5}{l}{$^d$ Triplet D$_{\infty h}$ symmetry}	
\end{tabular} 					
\end{table*}   

\begin{table*}[h!]
\centering
\caption{Total energies (in \ce{E_h}), equilibrium distance (in Bohr) and force constant (in a.u.) of triatomic systems for \ce{D$_{3h}$} symmetry calculated at VMC level }
\begin{tabular}{llll}
\hline
\textbf{System}        & {\textbf{E}}                   & \textbf{r$_{eq}$} & \textbf{$k_{\nu_1}$}  \\ \hline
\ce{2e^-[H^{3+}_3]} S  & -1.347(4)                      & 1.64(3)                               & 0.56(6)                          \\
\ce{2e^-[Li^{3+}_3]} S & -22.3345(2)                    & 5.687(10)                             & 0.0248(4)                        \\
\ce{2e^+[H^{3-}_3]} S  & -2.1489(1)                     & 6.149(10)                             & 0.0110(4)                        \\
\ce{2e^-[Li^{3+}_3]} T & -22.3105(2)                    & 5.950(7)                              & 0.0175(3)                        \\
\ce{2e^+[H^{3-}_3]} T  & -2.1266(1)                     & 6.78(2)                               & 0.0085(6)                        \\ \hline
\end{tabular}
\end{table*}

\begin{table*}[h!]
\centering
\caption{Total energies (in \ce{E_h}), equilibrium distance (in Bohr) and force constant (in a.u.) of triatomic systems for \ce{D$_{3h}$} symmetry calculated at DMC level. }
\begin{tabular}{lllll}
\hline
\textbf{System}        & {\textbf{E}}                   & \textbf{r$_{eq}$} & \textbf{$k_{\nu_1}$} & \textbf{$k_{\nu_2}$}  \\ \hline
\ce{2e^-[H^{3+}_3]} S  & -1.346(4)                      & 1.63(3)                               & 0.57(5)                                  & 0.16(2)                                  \\
\ce{2e^-[Li^{3+}_3]} S & -22.34190(10)                  & 5.645(7)                              & 0.0254(3)                                & 0.0110(4)                                \\
\ce{2e^+[H^{3-}_3]} S  & -2.1652(2)                     & 6.11(1)                               & 0.0114(4)                                & 0.0047(7)                                \\
\ce{2e^-[Li^{3+}_3]} T & -22.31280(7)                   & 5.937(7)                              & 0.0168(2)                                & -                                        \\
\ce{2e^+[H^{3-}_3]} T  & -2.1401(4)                     & 6.62(1)                               & 0.0080(4)                                & -                                        \\ \hline
\end{tabular}
\end{table*}

\begin{table*}[h!]
\centering
\caption{Potential energy curve (in Hartree) for \ce{2e^{+}[H^{2-}_3]} in \ce{D$_{3h}$} symmetry }
\begin{tabular}{lllll}
\hline
\multicolumn{1}{c}{\textbf{R/Bohr}} & \multicolumn{1}{c}{\textbf{VMC}} & \multicolumn{1}{c}{\textbf{$\pm$}} & \multicolumn{1}{c}{\textbf{DMC}} & \multicolumn{1}{c}{\textbf{$\pm$}} \\ \hline
4.4 & -2.060273 & 3.10E-04 & -2.084490 & 1.46E-04 \\
4.8 & -2.070135 & 1.08E-04 & -2.089042 & 1.38E-04 \\
5.2 & -2.072389 & 1.60E-04 & -2.091617 & 1.40E-04 \\
5.6 & -2.074593 & 1.15E-04 & -2.092395 & 1.52E-04 \\
6.0 & -2.075544 & 9.12E-05 & -2.092527 & 1.77E-04 \\
6.4 & -2.075353 & 8.09E-05 & -2.091662 & 2.24E-04 \\
6.8 & -2.074855 & 7.91E-05 & -2.090231 & 1.24E-04 \\
7.2 & -2.073976 & 9.49E-05 & -2.088270 & 1.49E-04 \\ \hline
\end{tabular}
\end{table*}

\begin{table*}[h!]
\centering
\caption{Potential energy curve (in Hartree) for \ce{e^{+}[H^{-}_3]} in \ce{D$_{3h}$} symmetry }
\begin{tabular}{lllll}
\hline
\multicolumn{1}{c}{\textbf{R/Bohr}} & \multicolumn{1}{c}{\textbf{VMC}} & \multicolumn{1}{c}{\textbf{$\pm$}} & \multicolumn{1}{c}{\textbf{DMC}} & \multicolumn{1}{c}{\textbf{$\pm$}} \\ \hline
4.4 & -1.752080 & 3.14E-04 & -1.777169 & 1.12E-04 \\
4.8 & -1.766600 & 1.13E-04 & -1.780472 & 9.55E-05 \\
5.2 & -1.770521 & 1.19E-04 & -1.782728 & 7.86E-05 \\
5.6 & -1.773680 & 1.08E-04 & -1.784643 & 9.21E-05 \\
6.0 & -1.775935 & 1.69E-04 & -1.786088 & 6.02E-05 \\
6.4 & -1.778397 & 1.18E-04 & -1.786910 & 6.99E-05 \\
6.8 & -1.779746 & 1.03E-04 & -1.787773 & 6.86E-05 \\
7.2 & -1.781522 & 1.00E-04 & -1.788336 & 5.98E-05 \\ \hline
\end{tabular}
\end{table*}

\begin{table*}[h!]
\centering
\caption{Potential energy curve (in Hartree) for \ce{e^{+}[H^{2-}_3]} in \ce{D$_{3h}$} symmetry }
\begin{tabular}{lllll}
\hline
\multicolumn{1}{c}{\textbf{R/Bohr}} & \multicolumn{1}{c}{\textbf{VMC}} & \multicolumn{1}{c}{\textbf{$\pm$}} & \multicolumn{1}{c}{\textbf{DMC}} & \multicolumn{1}{c}{\textbf{$\pm$}} \\ \hline
4.4 & -1.823325 & 2.64E-04 & -1.839220 & 1.22E-04 \\
4.8 & -1.828656 & 8.75E-05 & -1.843064 & 8.53E-05 \\
5.2 & -1.832172 & 7.46E-05 & -1.845263 & 1.10E-04 \\
5.6 & -1.833944 & 1.07E-04 & -1.846903 & 9.25E-05 \\
6.0 & -1.834647 & 8.87E-05 & -1.847079 & 1.02E-04 \\
6.4 & -1.834363 & 8.24E-05 & -1.846709 & 9.50E-05 \\
6.8 & -1.833510 & 7.79E-05 & -1.845503 & 1.13E-04 \\
7.2 & -1.831198 & 1.54E-04 & -1.844356 & 9.08E-05 \\ \hline
\end{tabular}
\end{table*}

\begin{table*}[h!]
\centering
\caption{Potential energy curve (in Hartree) for \ce{H^{-}_3} in \ce{D$_{3h}$} symmetry }
\begin{tabular}{lllll}
\hline
\multicolumn{1}{c}{\textbf{R/Bohr}} & \multicolumn{1}{c}{\textbf{VMC}} & \multicolumn{1}{c}{\textbf{$\pm$}} & \multicolumn{1}{c}{\textbf{DMC}} & \multicolumn{1}{c}{\textbf{$\pm$}} \\ \hline
4.8             & -1.540921    & 6.39E-05       & -1.548103    & 7.02E-05       \\
5.2             & -1.539070    & 7.60E-05       & -1.545813    & 7.25E-05       \\
5.6             & -1.537844    & 8.19E-05       & -1.544120    & 6.45E-05       \\
6.0             & -1.536943    & 6.66E-05       & -1.542198    & 5.99E-05       \\
6.4             & -1.535647    & 6.70E-05       & -1.540654    & 6.30E-05       \\
6.8             & -1.534728    & 5.68E-05       & -1.539143    & 5.01E-05       \\
7.2             & -1.533880    & 5.94E-05       & -1.537910    & 4.50E-05       \\ \hline
\end{tabular}
\end{table*}

\begin{table*}[h!]
\centering
\caption{Potential energy curve (in Hartree) for the positronic singlet \ce{2e^{+}[H^{3-}_3]} in \ce{D$_{3h}$} symmetry }
\begin{tabular}{lllll}
\hline
\multicolumn{1}{c}{\textbf{R/Bohr}} & \multicolumn{1}{c}{\textbf{VMC}} & \multicolumn{1}{c}{\textbf{$\pm$}} & \multicolumn{1}{c}{\textbf{DMC}} & \multicolumn{1}{c}{\textbf{$\pm$}} \\ \hline
4.4 & -2.124223 & 1.14E-04 & -2.143440 & 1.23E-04 \\
4.8 & -2.135587 & 1.16E-04 & -2.153232 & 1.47E-04 \\
5.2 & -2.142887 & 1.15E-04 & -2.159849 & 1.44E-04 \\
5.6 & -2.146867 & 1.15E-04 & -2.163421 & 1.67E-04 \\
6.0 & -2.148783 & 1.12E-04 & -2.165047 & 1.34E-04 \\
6.4 & -2.148563 & 1.12E-04 & -2.164726 & 9.51E-05 \\
6.8 & -2.146737 & 1.09E-04 & -2.162819 & 1.31E-04 \\
7.2 & -2.143985 & 1.07E-04 & -2.159856 & 1.46E-04 \\
7.6 & -2.140169 & 1.01E-04 & -2.156644 & 1.35E-04 \\ \hline
\end{tabular}
\end{table*}

\begin{table*}[h!]
\centering
\caption{Potential energy curve (in Hartree) for the positronic triplet \ce{2e^{+}[H^{3-}_3]} in \ce{D$_{3h}$} symmetry }
\begin{tabular}{lllll}
\hline
\multicolumn{1}{c}{\textbf{R/Bohr}} & \multicolumn{1}{c}{\textbf{VMC}} & \multicolumn{1}{c}{\textbf{$\pm$}} & \multicolumn{1}{c}{\textbf{DMC}} & \multicolumn{1}{c}{\textbf{$\pm$}} \\ \hline
5.2          & -2.113935          & 1.10E-04          & -2.129561          & 2.15E-04          \\ 
5.6          & -2.119879          & 9.09E-05          & -2.134930          & 1.69E-04          \\
6.0          & -2.123521          & 9.99E-05          & -2.138353          & 2.14E-04          \\
6.4          & -2.125958          & 1.04E-04          & -2.139902          & 3.38E-04          \\
6.8          & -2.126640          & 7.43E-05          & -2.139849          & 3.52E-04          \\
7.2          & -2.125796          & 8.89E-05          & -2.139030          & 4.96E-04          \\
7.6          & -2.124226          & 9.50E-05          & -2.137341          & 5.59E-04          \\
8.0          & -2.122099          & 9.37E-05          & -2.135430          & 6.40E-04          \\ \hline
\end{tabular}
\end{table*}

\begin{table*}[h!]
\centering
\caption{Potential energy curve (in Hartree) for the electronic singlet \ce{2e^{-}[Li^{3+}_3]} in \ce{D$_{3h}$} symmetry }
\begin{tabular}{lllll}
\hline
\multicolumn{1}{c}{\textbf{R/Bohr}} & \multicolumn{1}{c}{\textbf{VMC}} & \multicolumn{1}{c}{\textbf{$\pm$}} & \multicolumn{1}{c}{\textbf{DMC}} & \multicolumn{1}{c}{\textbf{$\pm$}} \\ \hline
4.8             & -22.320278   & 2.47E-04       & -22.329433   & 7.25E-05       \\
5.2             & -22.330751   & 2.14E-04       & -22.338943   & 5.71E-05       \\
5.6             & -22.334513   & 1.71E-04       & -22.341941   & 6.80E-05       \\
6.0             & -22.333354   & 3.16E-04       & -22.340373   & 6.50E-05       \\
6.4             & -22.329877   & 2.25E-04       & -22.336641   & 6.15E-05       \\
6.8             & -22.324713   & 2.42E-04       & -22.331058   & 7.34E-05       \\
7.2             & -22.318883   & 2.38E-04       & -22.324738   & 6.36E-05       \\
7.6             & -22.311375   & 2.12E-04       & -22.317742   & 4.81E-05       \\ \hline
\end{tabular}
\end{table*}

\begin{table*}[h!]
\centering
\caption{Potential energy curve (in Hartree) for the electronic triplet \ce{2e^{-}[Li^{3+}_3]} in \ce{D$_{3h}$} symmetry }
\begin{tabular}{lllll}
\hline
\multicolumn{1}{c}{\textbf{R/Bohr}} & \multicolumn{1}{c}{\textbf{VMC}} & \multicolumn{1}{c}{\textbf{$\pm$}} & \multicolumn{1}{c}{\textbf{DMC}} & \multicolumn{1}{c}{\textbf{$\pm$}} \\ \hline
4.8             & -22.294029   & 2.68E-04       & -22.296798   & 3.50E-05       \\
5.2             & -22.304349   & 1.80E-04       & -22.306955   & 3.38E-05       \\
5.6             & -22.309206   & 2.18E-04       & -22.311635   & 3.50E-05       \\
6.0             & -22.310410   & 2.18E-04       & -22.312654   & 3.70E-05       \\
6.4             & -22.309026   & 2.01E-04       & -22.311312   & 3.37E-05       \\
6.8             & -22.305786   & 2.38E-04       & -22.308146   & 3.33E-05       \\
7.2             & -22.301491   & 2.25E-04       & -22.304029   & 3.25E-05       \\
7.6             & -22.297133   & 1.79E-04       & -22.299412   & 3.10E-05       \\ \hline
\end{tabular}
\end{table*}

\begin{table*}[h!]
\centering
\caption{Potential energy curve (in Hartree) for \ce{H^{+}_3} in \ce{D$_{3h}$} symmetry }
\begin{tabular}{lllll}
\hline
\multicolumn{1}{c}{\textbf{R/Bohr}} & \multicolumn{1}{c}{\textbf{VMC}} & \multicolumn{1}{c}{\textbf{$\pm$}} & \multicolumn{1}{c}{\textbf{DMC}} & \multicolumn{1}{c}{\textbf{$\pm$}} \\ \hline
0.8             & -0.821350    & 6.89E-05       & -0.821701    & 1.19E-05       \\
0.9             & -1.004019    & 6.56E-05       & -1.004152    & 3.73E-06       \\
1.0             & -1.127085    & 7.61E-05       & -1.127251    & 1.13E-05       \\
1.1             & -1.210251    & 5.78E-05       & -1.210496    & 3.62E-06       \\
1.2             & -1.265961    & 5.00E-05       & -1.266223    & 1.44E-05       \\
1.4             & -1.325071    & 4.21E-05       & -1.325250    & 2.94E-06       \\
1.6             & -1.343060    & 3.87E-05       & -1.343246    & 1.04E-05       \\
1.8             & -1.339360    & 3.94E-05       & -1.339473    & 2.98E-06       \\
2.0             & -1.323952    & 3.23E-05       & -1.324112    & 1.06E-05       \\
2.2             & -1.302710    & 3.18E-05       & -1.302802    & 2.79E-06       \\
2.4             & -1.278603    & 2.56E-05       & -1.278797    & 1.03E-05       \\
2.6             & -1.253832    & 2.43E-05       & -1.253946    & 2.21E-06       \\
2.8             & -1.229235    & 2.67E-05       & -1.229398    & 9.39E-06       \\
3.2             & -1.183429    & 2.45E-05       & -1.183601    & 9.93E-06       \\
4.8             & -1.062555    & 1.75E-05       & -1.063170    & 2.03E-05       \\
5.2             & -1.046507    & 2.16E-05       & -1.046961    & 1.53E-05       \\
5.6             & -1.034108    & 2.07E-05       & -1.034581    & 1.45E-05       \\
6.0             & -1.024797    & 2.51E-05       & -1.025332    & 1.60E-05       \\
6.4             & -1.017976    & 2.29E-05       & -1.018502    & 1.57E-05       \\
6.8             & -1.013012    & 1.84E-05       & -1.013440    & 2.06E-05       \\
7.2             & -1.009327    & 2.40E-05       & -1.009663    & 2.03E-05       \\ \hline
\end{tabular}
\end{table*}

\begin{table*}[h!]
\centering
\caption{ Potential energy curve (in Hartree) for \ce{2e^{+}[H^{2-}_3]} in \ce{C$_{2v}$} symmetry as a function of the internal angle $\theta$ (in degrees) while fixing two of the internuclear distances to R=6.1 bohr  }
\begin{tabular}{ccllll}
\hline
\textbf{Angle $\theta$} & \textbf{$R_3$} & \multicolumn{1}{c}{\textbf{VMC}} & \multicolumn{1}{c}{\textbf{$\pm$}} & \multicolumn{1}{c}{\textbf{DMC}} & \multicolumn{1}{c}{\textbf{$\pm$}} \\ \hline
20 & 2.14 & -2.152772 & 3.90E-04 & -2.162633 & 1.02E-03 \\
25 & 2.66 & -2.090512 & 2.89E-04 & -2.115209 & 2.66E-04 \\
30 & 3.18 & -2.079752 & 2.85E-04 & -2.101476 & 1.49E-04 \\
35 & 3.70 & -2.047689 & 8.17E-05 & -2.096023 & 4.00E-04 \\
40 & 4.21 & -2.032580 & 9.46E-05 & -2.092927 & 3.01E-04 \\
50 & 5.20 & -2.022538 & 1.16E-04 & -2.089545 & 3.18E-04 \\
60 & 6.15 & -2.018736 & 9.51E-05 & -2.087508 & 3.19E-04 \\
70 & 7.05 & -2.018641 & 1.66E-04 & -2.087092 & 3.22E-04 \\
80 & 7.91 & -2.018582 & 1.33E-04 & -2.086885 & 2.49E-04 \\ \hline
\end{tabular}
\end{table*}

\begin{table*}[h!]
\centering
\caption{ Potential energy curve (in Hartree) for \ce{e^{+}[H^{2-}_3]} in \ce{C$_{2v}$} symmetry as a function of the internal angle $\theta$ (in degrees) while fixing two of the internuclear distances to R=6.1 bohr  }
\begin{tabular}{ccllll}
\hline
\textbf{Angle $\theta$} & \textbf{$R_3$} & \textbf{VMC} & \textbf{$\pm$} & \textbf{DMC} & \textbf{$\pm$} \\ \hline
20 & 2.14 & -1.906276 & 3.60E-04 & -1.910344 & 2.43E-04 \\
25 & 2.66 & -1.862379 & 6.80E-05 & -1.873857 & 8.22E-05 \\
30 & 3.18 & -1.852850 & 8.93E-05 & -1.863797 & 9.40E-05 \\
35 & 3.70 & -1.847789 & 1.08E-04 & -1.859888 & 8.31E-05 \\
40 & 4.21 & -1.845121 & 7.63E-05 & -1.856901 & 1.04E-04 \\
50 & 5.20 & -1.838761 & 1.32E-04 & -1.851839 & 9.61E-05 \\
60 & 6.15 & -1.834154 & 1.22E-04 & -1.847309 & 1.02E-04 \\
70 & 7.05 & -1.835532 & 9.99E-05 & -1.846307 & 1.10E-04 \\
80 & 7.91 & -1.835666 & 1.04E-04 & -1.847058 & 9.57E-05 \\ \hline
\end{tabular}
\end{table*}

\begin{table*}[h!]
\centering
\caption{ Potential energy curve (in Hartree) for \ce{e^{+}[H^{-}_3]} in \ce{C$_{2v}$} symmetry as a function of the internal angle $\theta$ (in degrees) while fixing two of the internuclear distances to R=6.1 bohr  }
\begin{tabular}{ccllll}
\hline
\textbf{Angle $\theta$} & \textbf{$R_3$} & \textbf{VMC} & \textbf{$\pm$} & \textbf{DMC} & \textbf{$\pm$} \\ \hline
20 & 2.14 & -1.908521 & 7.24E-05 & -1.913757 & 5.16E-05 \\
25 & 2.66 & -1.862696 & 7.40E-05 & -1.868503 & 6.98E-05 \\
30 & 3.18 & -1.799373 & 1.73E-04 & -1.828375 & 1.23E-04 \\
35 & 3.70 & -1.803534 & 8.20E-05 & -1.811682 & 6.27E-05 \\
40 & 4.21 & -1.790276 & 8.11E-05 & -1.800972 & 7.85E-05 \\
50 & 5.20 & -1.782268 & 2.09E-04 & -1.794755 & 1.99E-04 \\
60 & 6.15 & -1.783209 & 1.34E-04 & -1.793528 & 2.62E-04 \\
70 & 7.05 & -1.781331 & 3.33E-04 & -1.792499 & 1.73E-04 \\
80 & 7.91 & -1.778940 & 8.52E-05 & -1.793478 & 1.97E-04 \\ \hline
\end{tabular}
\end{table*}

\begin{table*}[h!]
\centering
\caption{ Potential energy curve (in Hartree) for \ce{[H^{-}_3]} in \ce{C$_{2v}$} symmetry as a function of the internal angle $\theta$ (in degrees) while fixing two of the internuclear distances to R=6.1 bohr  }
\begin{tabular}{ccllll}
\hline
\textbf{Angle $\theta$} & \textbf{$R_3$} & \textbf{VMC} & \textbf{$\pm$} & \textbf{DMC} & \textbf{$\pm$} \\ \hline
20 & 2.14 & -1.649986 & 6.38E-05 & -1.651792 & 2.89E-05 \\
25 & 2.66 & -1.604368 & 4.96E-05 & -1.606617 & 3.39E-05 \\
30 & 3.18 & -1.568879 & 5.86E-05 & -1.572085 & 4.50E-05 \\
35 & 3.70 & -1.546772 & 1.30E-04 & -1.556130 & 5.95E-05 \\
40 & 4.21 & -1.543869 & 7.77E-05 & -1.548743 & 5.20E-05 \\
50 & 5.20 & -1.538636 & 7.83E-05 & -1.544087 & 6.23E-05 \\
60 & 6.15 & -1.535500 & 5.76E-05 & -1.541480 & 8.02E-05 \\
70 & 7.05 & -1.535484 & 6.07E-05 & -1.540173 & 6.09E-05 \\
80 & 7.91 & -1.534719 & 1.48E-04 & -1.541791 & 7.13E-05 \\ \hline
\end{tabular}
\end{table*}

\begin{table*}[h!]
\centering
\caption{ Potential energy curve (in Hartree) for \ce{2e^{+}[H^{3-}_3]} in \ce{C$_{2v}$} symmetry as a function of the internal angle $\theta$ (in degrees) while fixing two of the internuclear distances to R=6.1 bohr  }
\begin{tabular}{llllll}
\hline
\textbf{Angle $\theta$} & \textbf{$R_3$} & \textbf{VMC} & \textbf{$\pm$} & \textbf{DMC} & \textbf{$\pm$} \\ \hline
20 & 2.14 & -2.138950 & 1.26E-04 & -2.174117 & 5.44E-04 \\
25 & 2.66 & -2.114563 & 1.17E-04 & -2.141195 & 2.46E-04 \\
30 & 3.18 & -2.117330 & 1.13E-04 & -2.138720 & 1.71E-04 \\
35 & 3.70 & -2.126270 & 1.20E-04 & -2.145415 & 1.36E-04 \\
40 & 4.21 & -2.135539 & 9.17E-05 & -2.153423 & 1.26E-04 \\
50 & 5.20 & -2.145800 & 9.90E-05 & -2.163058 & 1.17E-04 \\
60 & 6.15 & -2.149130 & 1.22E-04 & -2.165092 & 1.12E-04 \\
70 & 7.05 & -2.146746 & 1.41E-04 & -2.163734 & 1.23E-04 \\
80 & 7.91 & -2.143824 & 8.35E-05 & -2.160592 & 1.25E-04 \\ \hline
\end{tabular}
\end{table*}

\InputIfFileExists{bibliography.bbl}{}{}



\end{document}